\documentclass[manuscript]{acmart}
\AtBeginDocument{%
  \providecommand\BibTeX{{%
    \normalfont B\kern-0.5em{\scshape i\kern-0.25em b}\kern-0.8em\TeX}}}

\setcopyright{acmcopyright}
\copyrightyear{2018}
\acmYear{2018}
\acmDOI{10.1145/1122445.1122456}

\acmConference[Woodstock '18]{Woodstock '18: ACM Conference on Human-Computer Interaction}{June 03--05, 2018}{Woodstock, NY}
\acmBooktitle{Woodstock '18: ACM Conference on Human-Computer Interaction, June 03--05, 2018, Woodstock, NY}
\acmPrice{15.00}
\acmISBN{978-1-4503-XXXX-X/18/06}

\usepackage{tabularx}
\usepackage{enumitem}
\usepackage[toc,page]{appendix}
\usepackage{multirow}
\usepackage{makecell}
\usepackage{hhline}
\usepackage{afterpage}

\begin{document}


\title{``I never would have thought to say this'': Example-Based Exploration to Balance Scientists' Writing Preferences with Public Science Communication Strategies}
% \shorttitle{Audience Impressions of Informal Science Communication}
\renewcommand{\shorttitle}{Balancing Scientists' Writing Preferences with Science Communication Strategies}

% \author{Anonymous For Submission}
\author{Grace Li}
\email{gl2676@columbia.edu}
\affiliation{
    \institution{Columbia University}
    \city{NY}
    \state{NY}
    \country{USA}
}

\author{Yuanyang ``YY'' Teng}
\affiliation{
    \institution{Columbia University}
    \city{NY}
    \state{NY}
    \country{USA}
}

\author{Juna Kawai-Yue}
\affiliation{
    \institution{Columbia University}
    \city{NY}
    \state{NY}
    \country{USA}
}

\author{Unaisah Ahmed}
\affiliation{
    \institution{Barnard College}
    \city{NY}
    \state{NY}
    \country{USA}
}

\author{Anatta S. Tantiwongse}
\affiliation{
    \institution{Columbia University}
    \city{NY}
    \state{NY}
    \country{USA}
}

\author{Jessica Y. Liang}
\affiliation{
    \institution{Columbia University}
    \city{NY}
    \state{NY}
    \country{USA}
}

\author{Dorothy Zhang}
\affiliation{
    \institution{Columbia University}
    \city{NY}
    \state{NY}
    \country{USA}
}

\author{Kynnedy Simone Smith}
\affiliation{
    \institution{Columbia University}
    \city{NY}
    \state{NY}
    \country{USA}
}

\author{Tao Long}
\affiliation{
    \institution{Columbia University}
    \city{NY}
    \state{NY}
    \country{USA}
}

\author{Mina Lee}
\affiliation{
    \institution{University of Chicago}
    \city{Chicago}
    \state{IL}
    \country{USA}
}

\author{Lydia B Chilton}
\affiliation{
    \institution{Columbia University}
    \city{NY}
    \state{NY}
    \country{USA}
}
% \affiliation{%


\newcommand{\yy}[1]{{\color{purple}\bf{YY Comments: #1}\normalfont}}
\newcommand{\grace}[1]{{\color{pink}\bf{Grace Comments: #1}\normalfont}}
\newcommand{\lydia}[1]{{\color{red}\bf{Lydia Comments: #1}\normalfont}}
\def \revision #1{\textcolor{blue}{#1}}
%TC:ignore
\begin{abstract}

Public-facing science communication is important in garnering interest, engagement, and trust in science. Social media platforms provide scientists with opportunities to reach broader audiences, yet many resist adopting social media writing strategies because the strategies conflict with traditional science writing norms and personal preferences. To address this gap, we first evaluate readers’ preferences for strategies such as examples, walkthroughs, and personal language. While many readers enjoyed science narratives that used these strategies, their effectiveness was nuanced and context-dependent, varying by topic and individual preference. Building on these findings, we design a system that uses contrastive examples to help scientists adopt and integrate these social media science writing strategies. In a user study with scientists, we found that presenting contrastive examples helped writers critically evaluate different narrative options, balance competing goals, and gain confidence in adapting social media writing strategies to fit both their topic and audience.

\end{abstract}
%TC:endignore

% lydia 
% Audience study 
% Writer's Study





\maketitle

\section{Introduction}
Scientists go through years of training to learn the structure and style of writing research papers as a way for other scientists in their field to learn about their work \cite{pearson2001participation, powell2008building}. But sharing science beyond these formal research communities is important in supporting public interest in science \cite{national2017communicating, burns2003science}. Science communication on social media platforms, like X, has emerged as a straightforward and accessible way for scientists to communicate with the public \cite{hou2017hacking, williams2022hci, bruggemann2020post}. To do so, scientists must learn a different set of writing structures and styles to use when communicating science to an everyday audience on social media \cite{gilbert2020run, della2021expert}. 

Prior work has identified specific strategies of successful science posts around how to connect with an everyday audience via relatable examples, a step-by-step walkthrough, and personal language \cite{gero2021makes}. But this type of writing is often at odds with how scientists were trained to write about science. Prior work has found that scientists struggle to negotiate their identities as scientists with the expectations of social media \cite{lorono2018responsibility}. For example, scientists might question whether using personal language will detract from their credibility as a scientist \cite{ 10.1145/3479566, koivumaki2020social}. 

In order to support scientists in engaging with this type of science communication, we need to understand readers’ preferences of these different writing strategies to help scientists feel more comfortable adopting social media writing strategies. To do so, we evaluate readers’ perspectives on these different social media science communication strategies: 1) use of a relatable example – such as using the iridescence of bubbles to motivate the topic “Thin Film Interference" in physics, 2) use of a step-by-step walkthrough that uses 10-12 tweets which use an example to explain every step, and 3) use of personal language – the writer talking about themselves and their experience, often in a personal and conversational way, “Crazy, right? WTH was I thinking ?!?” \cite{gero2021makes}. Prior research shows that teaching writers about these strategies doesn’t help them incorporate them into their writing for social media \cite{gero2021makes}. Using the findings from the reader study, we design a system that provides scientists with different examples of the science communication strategies to help scientists balance their own science writing background with these social media writing strategies.

Overall, readers did prefer the use of a relatable example and personal language, but were split for the use of a walkthrough. While many readers enjoyed the use of personal language to make the science narratives more engaging, there were readers who found personal language to be distracting to the science. Similarly, while many readers preferred examples, some readers found that the use of an example often added unnecessary details to explaining science or isolated readers because the example used was unrelatable. Because of readers' differing preferences, these writing strategies do not work for all readers and their effectiveness varies based on the topic. 

Given the diverse and contextual preferences that readers had for
these science communication strategies, static writing guidelines do not accommodate these preferences. We design a system that presents scientists a set of options for how to (1) structure and (2) style their writing. We find that presenting each dimension of social media writing strategies through different options helps scientists balance their science background and readers’ preferences. Presenting different structure and style options allows writers to choose the degree to which they employ a given social media strategy, turning  static writing strategies into a spectrum of options for writers to adopt according to their own preferences.

We find that showing scientists options allows them to more easily adopt these science communication strategies while still maintaining their own preferences and beliefs about how science should be communicated. Instead of presenting science communication as an immutable set of guidelines, we find that presenting writers with different structure and style options can support scientists in making more informed and intentional decisions about when and how to integrate these social media writing strategies.

\section{Background on Tweetorials}
Tweetorials are a form of social media science communication on Twitter, defined as a chronological series of tweets that explains a science topic \cite{symplurTweetorialsFrom, Bruggemann2020-ez}. According to Breu and Berstein, Tweetorials emerged from the medical community to continue education for other medical professionals, often containing medical jargon and not intended for everyday audiences. As the form became popularized, other communities on Twitter began to adopt the format \cite{doi:10.1056/NEJMp1906790, symplurTweetorialsFrom, Bruggemann2020-ez}. As a result, Breu's definition of a Tweetorial, ``a collection of threaded tweets aimed at teaching users who engage with them,'' can be applied to various domains such as biology, computer science, and economics. 

Gero et. al. identified and compared key features of Tweetorials to traditional science communication strategies \cite{10.1145/3479566}. The authors identified 3 main structural components of Tweetorials to be the hook, body, and conclusion. Previous research has explored strategies for writing engaging hooks \cite{10.1145/3643834.3661587, long2023tweetorialhooksgenerativeai}. Thus, we focus on specific strategies that appear within the body of a Tweetorial. 

\subsection{Key Strategies of the Tweetorial Body}
According to Gero et. al., the body of a Tweetorial is the most varied in length and types of detail used to explain a given topic \cite{10.1145/3479566}. The researchers found that Tweetorials contain specific techniques such as the use of an example, a step-by-step walkthrough structure, and personal language. 

\subsubsection{Example (E)}
\label{example_defintion}
Like traditional science communication, Tweetorials often contain an example that uses familiar or simpler concepts to explain the main idea \cite{10.1145/3479566}. We define this as an \textbf{example (E)} technique. Components of the example are the \textit{use case} and the \textit{scenario}. The \textit{use case} is a general application of the scientific topic and the \textit{scenario} is a specific situation that describes how the science topic was applied. For example, one Tweetorial uses the example of fingertips getting wrinkly in the bath to explain water immersion wrinkling.\footnote{\href{http://language-play.com/tech-tweets/tweetorial/4}{http://language-play.com/tech-tweets/tweetorial/4}} For this Tweetorial, the use case is when fingers get wrinkly in water. The scenario is a parent bathing their child. We use the use case and scenario for more finegrain control over the LLM-generated narratives in Section \ref{narrative_generation_strategy}. 

% Another 

% \grace{where to add descirption fo data inputs}
% The [example] data field is in the format of a short phrase. For instance: \textit{``Spider’s Web”} is the [example] for the topic Tensile Structure in Civil Engineering. The [scenario] data field contains a scenario of a personal narrative that narrates the example through a scenario and connects with the scientific topic. For instance, for the \textit{``Spider’s Web”} example, the scenario is \textit{``During our late-night camping, my adventurous friend decided to challenge a playful spider. He picked up a small twig and started slowly poking its web. Upon noticing the twig, the spider ran towards it, displaying its territorial instinct. I witnessed how the web, a miraculous tensile structure, withheld the pressure without falling apart. Thanks to the constant tension in the silk material, the web stayed steady, absorbing the additional load, distributed throughout its double-curved surface, and transmitting it to its anchor points.”}. We used the prompt seen in Figure 

\subsubsection{Walkthrough (W)}
Tweetorial structures often use a narrative and signposting to establish a narrative structure. We define these two attributes as the \textbf{walkthrough (W)} technique. Narrative is defined by a series of connected events to explain a given topic. Tweetorials signpost by using transition words like ``Firstly'' and ``Secondly,'' or by using a list of questions to help frame the sequence of the Tweetorial. In a Tweetorial about selectivity metrics in college rankings, the author uses the second tweet to list out 3 driving questions for the explanation and to establish the structure of the Tweetorial: ``1. Does ``selectivity'' actually tell you anything useful about how good your education will be? 2. What does ``selectivity'' actually measure that is of value to a student? 3. Why do I have the feeling somebody chose this metric cause they just needed more stuff to rank by?''\footnote{\href{http://language-play.com/tech-tweets/tweetorial/14}{http://language-play.com/tech-tweets/tweetorial/14}} 
% \yy{the [] in the quote is easily confused with citations, and they are not in the actual tweetorial. This example is more aligned with motivation after the hook, rather than a body walkthrough}

\subsubsection{Personal Language (P)}
Tweetorials often use subjective, conversational, and informal language. We define these features as the \textbf{personal language (P)} technique.  Some authors might use first-person pronouns like ``I'' to talk from their subjective perspective,\footnote{\href{http://language-play.com/tech-tweets/tweetorial/1}{http://language-play.com/tech-tweets/tweetorial/1}} or use the second person, ``you,'' to directly address the audience in conversation: ``You can think of a Hash Function like a magic fingerprint reader.''\footnote{\href{http://language-play.com/tech-tweets/tweetorial/31}{http://language-play.com/tech-tweets/tweetorial/31}} Some authors might use ALLCAPS or emojis to engage in informal language and humor: ``OH MAN MY HEAD HURTS AND MY LIMBS TINGLE EVERY TIME I GO TO A CHINESE RESTAURANT, I THINK IT MAY BE ALL THE MSG THEY PUT IN THE FOOD???''.\footnote{\href{http://language-play.com/tech-tweets/tweetorial/33}{http://language-play.com/tech-tweets/tweetorial/33}} 

\begin{figure}
    \centering\includegraphics[width=0.62\linewidth]{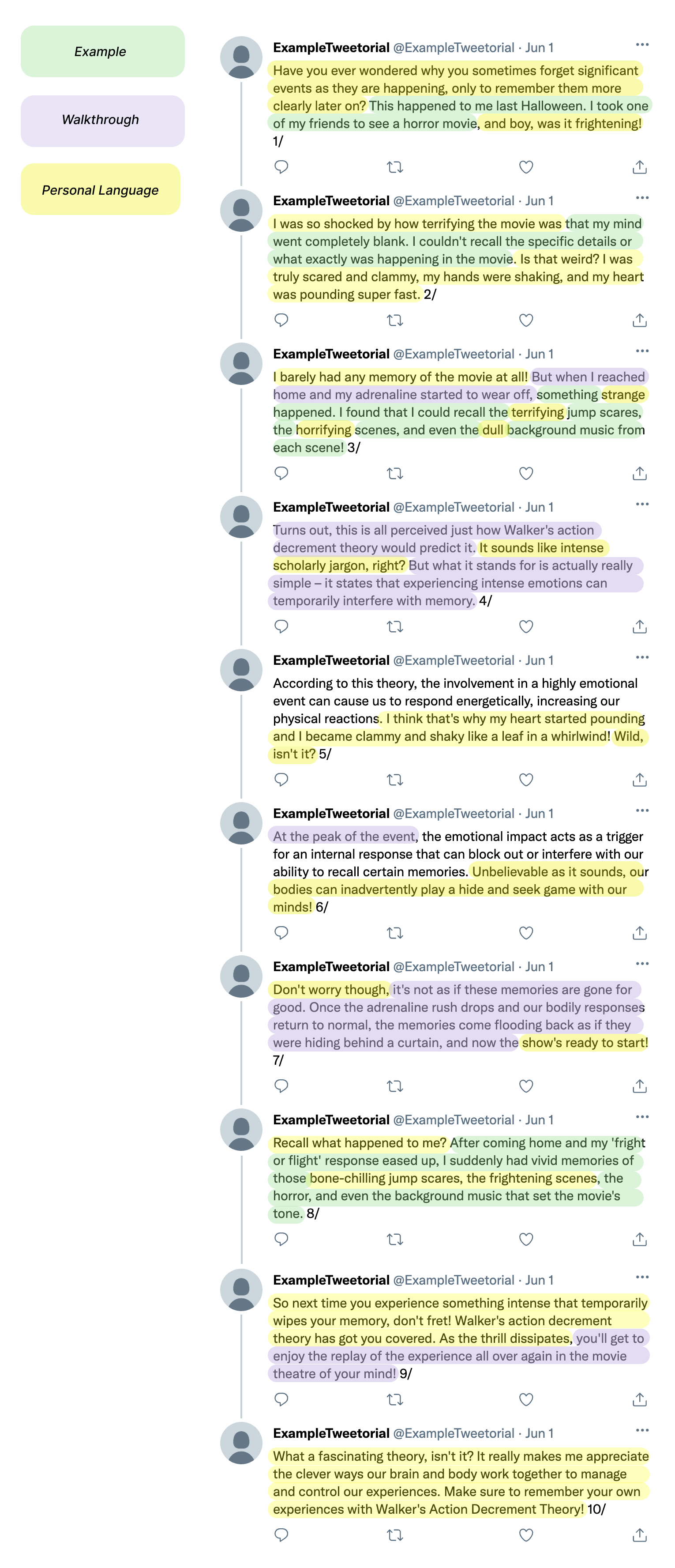}
    \caption{Annotated Tweetorial on the topic of Walker's Action Decrement Theory in Psychology with color highlights corresponding to Example, Walkthrough, and Personal Language.}
    \label{fig:everything}

\end{figure}

\vspace{5mm}
Figure \ref{fig:everything} provides an annotated Tweetorial highlighting these three techniques (example (E), walkthrough (W), and personal language (P)) on the topic of Walker's Action Decrement Theory in Psychology to demonstrate how they are applied in a science explanation. We use these three techniques to ground our approach in understanding reader preferences for science communication on social media and how writers explore the design space for science writing structures and styles.

\section{Related Work}

\subsection{HCI for Science Communication}
% \grace{add some more bits from, edit intro

% 1. motivate this more, doing a deeper indepth XYZ to see how audience respond to particular aspects of writing, generalize aspects of writing
% 2. related work defends what we do is novel [LLM section]}

% Science communication has transitioned away from traditional scholarly articles and towards more informal platforms such as blogs and social media \cite{hou2017hacking, williams2022hci}. 
% \grace{take paragraph 1, cut last sentence
% take paragraph 2, --> add to the end of paragraph 2, research find it hard ebcause of blah blah --> use these findings to support people }

Science communication is integral to engage everyday people in science advancements and help them understand the world around them. Research has explored how new forms of science communication, such as Tweetorials, are situated among other genres, how different social media platforms shape the expectations that readers have, readers' interaction behaviors with science writing, and the way readers access scientific information \cite{heyd2015digital, williams2022hci, hargittai2018young, tardy2023spread}. 

% recommendations for communication systematic review of paper

Science communication on social media is informal and often told from a personal perspective to engage a broader audience \cite{10.1145/3479566}. Previous work has compiled recommendations for effective science communication on social media based on how likely it is that using these strategies will increase engagement: comments, likes, and shares  \cite{zhang2022no, bik2015ten, fontaine2019communicating}. But these works do not consider whether these techniques are actually effective for readers to understand the science. Our work evaluates whether these different science communication strategies on Twitter are effective in helping readers engage with and understand the science.  

For scientists seeking to engage with science communication on social media, they struggle to understand their ``audience'' which consists of readers from diverse backgrounds, levels of scientific literacy, and preferences which makes it challenging for scientists to determine \textit{who} they are writing for \cite{rice2017contexts, schafer2017changing, williams2022hci}. Furthermore, each social media platform has its own set of community norms and demographics \cite{pewresearchFactsAbout, oktay2014demographic}. Research has also shown how scientists struggle to reach non-research audiences on social media and to adapt to these new genres of science communication \cite{cote2018scientists, kopke2019stepping, lorono2018responsibility, koivumaki2020social}. Even though language with strong sentiment can predict success on social media, many scientists are hesitant about such language having negative effects on their credibility \cite{yuan2020s, yuan2019should, gilbert2020run}. We use these findings to explore how providing different structure and style options to writers can help accommodate different writers' hesitancy and comfort levels with science communication on social media.

\subsection{LLM-based Writing Support}
Recent advances in AI writing capabilities have accelerated work in LLM-based writing interfaces \cite{biermann2022tool, buschek2021impact, shakeri2021saga, sommers1980revision, lee2024design}. Our contributions are most directly tied to the specific writing task of drafting (writing stage) science narrative explanations (purpose) for an everyday audience (audience) \cite{lee2024design}.

Previous research has explored how analogies can be used to translate technical content into more relatable references for people \cite{august2020writing, kim2024authors, nguyen2024simulating, hullman2018improving}. LLM-based systems have been used to explore different ways to generate engaging and relatable hooks for science narratives \cite{long2023tweetorialhooksgenerativeai, 10.1145/3643834.3661587}. LLMs have been used for the personalization of scientific information to improve comprehension, alignment to individual preferences, and accommodate different science literacy levels \cite{august2023paper, das2023balancing, ding2023fluid}. But these cases focus on generating relatable examples or translating technical jargon into everyday language and do not connect the corresponding examples to an overarching narrative to explain the science. 

LLMs have been used in narrative generation to support story writing by tailoring generated narratives to user inputs such as story elements, topics, or rough sketches of plot development \cite{calderwood2022spinning, belz2024story, park2023designing, rashkin2020plotmachines}. New methods have explored ways to improve coherence in LLM-generated long-form story content through recursive prompting and revision and outline control \cite{yang2022re3, yang2022doc, wang2023improving}. While advancements have been made in improving coherence in LLM-generated long-form stories, generalizing these methods to science narratives is hard. Science narratives need both a coherent story and an accurate representation of the science. Our work explores how LLMs can support scientists in drafting a science narrative. To do so, we bridge research in relatable example generation with story narrative generation to create science narratives that follow a cohesive narrative structure around a relatable example. We provide scientists a baseline science narrative to iterate on.

\section{Reader Study: Methodology}

\subsection{Research Questions}
Tweetorials typically include three techniques \textbf{example (E)}, \textbf{walkthrough (W)}, and \textbf{personal language (P)} \cite{10.1145/3479566}. We want to understand how each of these techniques affects readers' ratings of science communication on social media. Given that most published Tweetorials contain these three techniques, we hypothesize that science explanations that contain all three features, \textbf{EWP}, will have the highest reader preference rating. To evaluate the effect of the 3 different features, we compare narratives with all three features (\textbf{EWP}) to narratives with one of the features removed (\textbf{EW}, \textbf{EP}, \textbf{WP}). We investigate the following hypotheses in a survey study with readers:

\textbf{H1: Example (E)}: Readers prefer explanations with an example (EWP) compared to those without an example (WP). 

\textbf{H2: Walkthrough (W)}: Readers prefer explanations that include a step-by-step walkthrough of the topic using an example (EWP) compared to explanations with multiple unrelated examples (EP). 

\textbf{H3: Personal Language (P)}: 
Readers prefer explanations that use personal and subjective language (EWP) compared to explanations that have a neutral scientific voice (EW).

% \textbf{H4: All three techniques (EWP)}: Readers prefer explanations that use all three techniques of example, walkthrough, and personal language (EWP) compared to explanations that use a subset of the three techniques. \grace{we don't actually test these techniques invidiually, the accurate H4 would be: Readers prefer explanations with all 3 explanation techniques compared to explanations that only use 2 of the 3 techniques. (which i think is basically equivalent to finding H1, H2, H3...)}

Testing each of these hypotheses requires comparing two explanations for the same scientific topic that are as similar as possible and only vary in whether they contain a example, walkthrough or personal language. Parallel examples like this are unlikely to occur naturally. Thus, we use AI to generate parallel explanations in each condition (See Section \ref{narrative_generation_strategy}). 

\color{black}

\subsection{Automatic Narrative Generation Strategy}
\label{narrative_generation_strategy}

We used OpenAI's GPT-4 API to generate science explanations with and without each technique in the form of Tweetorials (about 10 tweets in length) to ensure consistency and to make fair comparisons between science explanations. We describe the method for each technique in Sections \ref{H1_Prompting}, \ref{H2_Prompting}, and \ref{H3_Prompting}. 

We cover 5 diverse STEM fields: a physical science field (Physics), a social science field (Psychology), a technological field (Computer Science), a mathematical field (Statistics), and an engineering field (Civil Engineering). For each field an expert selected topics for 3 different levels of complexity (introductory, intermediate, and advanced levels) for a total of 15 topics (Appendix \ref{stem_topics}). We generated 75 different science explanations (5 conditions for each topic) for readers to rate. Each science explanation was validated by a corresponding expert for accuracy. We describe exact prompting methods for each hypothesis (H1: Example (E), H2: Walkthrough (W), H3: Personal Language(P)) below.

\subsubsection{Generating Tweetorials with and without Examples (H1)} 
\label{H1_Prompting}

We used GPT-4 to generate science narratives that contain an \textbf{E}xample, \textbf{W}alkthrough, and \textbf{P}ersonal language. We included five few-shot examples of published Tweetorials on Twitter to create our experimental condition: \textbf{EWP}. Experts on each topic provided data inputs [use case] and [scenario] (described in Section \ref{example_defintion}) to define the specific example the given narrative should use throughout the explanation. We provided specific guidelines regarding how the LLM should incorporate the given example, a walkthrough, and personal language. We iterated on each line separately and in compilation to ensure that the prompt was concise, essential, and reasonably consistent (Appendix \ref{prompts}). 

To generate the baseline condition, \textbf{WP}, we use a ``remove" method which provides GPT-4 a given narrative and a set of guidelines that specifies what technique to remove from the given narrative while maintaining all other conditions. The output is a new narrative with only the specified technique removed. To generate a science narrative without an example, we use the ``remove" method on the example. We pass in the experimental condition narration, \textbf{EWP}, and a set of guidelines that specify only the example should be removed from the narrative while maintaining all other elements such as structure and style (Appendix \ref{WP_prompt}). We use this same prompting strategy to ``remove personal language" in Section \ref{H3_Prompting}.

\begin{figure}
    \centering
    \includegraphics[width=0.9\linewidth]{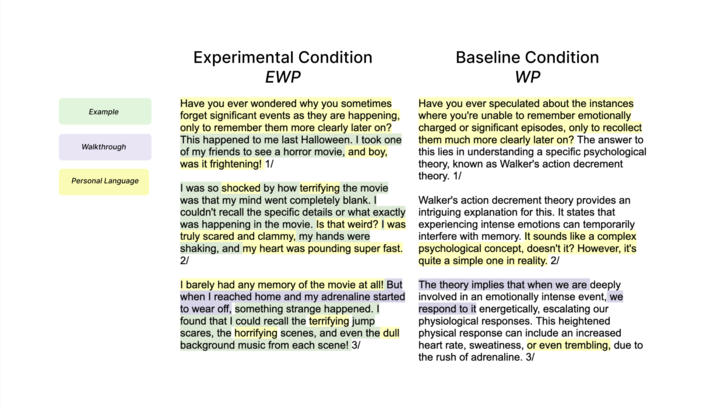}
    \caption{\textbf{H1: Example} Sample explanation generations on the topic of "Walker's Action Decrement Theory" in Psychology comparing the experimental condition (EWP) and baseline condition (WP).}
    \label{fig:example}
\end{figure}

Figure \ref{fig:example} shows a side-by-side annotated example of the experimental condition (EWP) and baseline condition (WP) for the topic of Walker's Action Decrement Theory. The lack of example highlights in green in the baseline condition shows the effect of no example in contrast with the explanation with everything.

\subsubsection{Generating Tweetorials with and without Personal Language (H3)} \label{H3_Prompting}
To understand reader preferences for science explanations with and without personal language (H2:Personal Language), we followed a similar GPT-4 generation protocol as in Section \ref{H1_Prompting} to generate the experimental condition which contains all three techniques and few-shot examples, \textbf{EWP}. To generate the baseline condition, we used the ``remove" procedure from Section \ref{H1_Prompting} to ``remove personal language" from the experimental condition, \textit{EWP}. We pass in \textit{EWP} and specify guidelines to only remove the personal language while maintaining all structural elements of the science explanation to create \textbf{EW}. 

\begin{figure}
    \centering
    \includegraphics[width=0.9\linewidth]{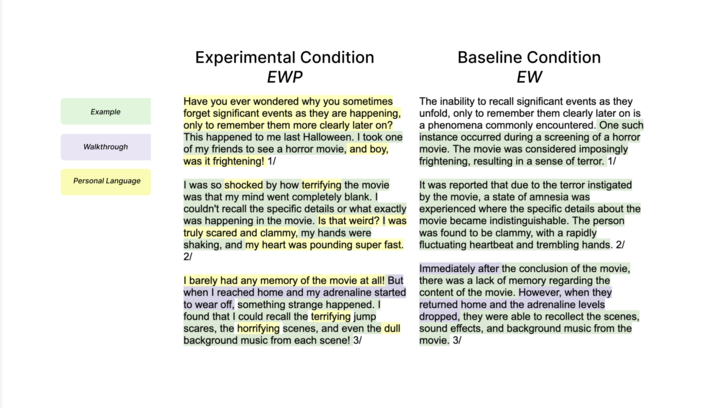}
    \caption{\textbf{H3:Personal Language} Sample explanation generations on the topic of Walker's Action Decrement Theory in Psychology comparing the experimental condition (EWP) and baseline condition (EW).}
    \label{fig:personal}
\end{figure}

Figure \ref{fig:personal} shows a side-by-side annotated example of the experimental condition (EWP) and baseline condition (EW) for the topic of Walker's Action Decrement Theory. The lack of personal language highlights in yellow in the baseline condition shows the effect of removing personal language from EWP. 

\subsubsection{Generating Tweetorials with and without Walkthroughs (H2)} \label{H2_Prompting}

To understand reader preferences for walkthroughs (H2:Walkthrough), we used GPT-4 to generate two different science explanations with contrasting narrative structures. In preliminary testing, we found that the ``remove" method failed to generate narratives without walkthroughs because the walkthrough served as the narrative structure. Thus, we provide the data inputs of [topic] and [domain] and a new set of guidelines to GPT-4 to generate narratives without a walkthrough (Appendix \ref{EP_prompt}). The guidelines specify instructions for not using a walkthrough (e.g. the explanation should use a non-sequential approach and that every tweet stands alone). No few-shot examples were added because there were no published Tweetorials with no walkthrough to create \textbf{EP-NoFewShot}.

\begin{figure}
    \centering
    \includegraphics[width=0.9\linewidth]{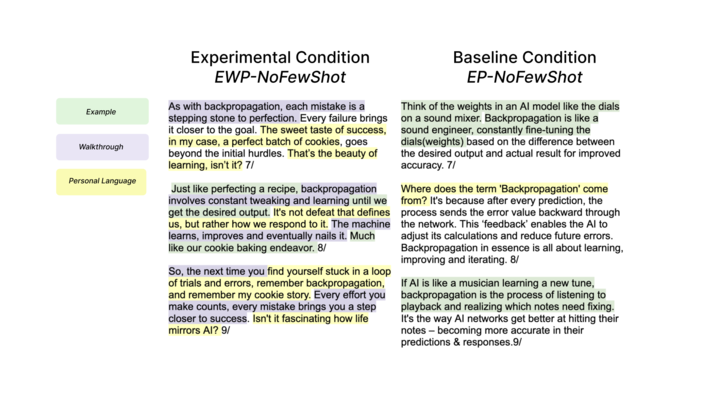}
    \caption{\textbf{H2: Walkthrough} Sample explanation generations on the topic of Back Propagation in Computer Science contrasting experimental condition (EWP-NoFewShot) and baseline condition (EP-NoFewShot).}
    \label{fig:walkthrough}
\end{figure}

We used the same base prompt as the experimental conditions in Section \ref{H1_Prompting} and Section \ref{H2_Prompting} to generate a science explanation with an \textbf{E}xample, \textbf{W}alkthrough and \textbf{P}ersonal language. To maintain a fair comparison with EP-NoFewShot, we omitted few-shot examples to create \textbf{EWP-NoFewShot}. Figure \ref{fig:walkthrough} highlights the differences between science narratives with and without walkthroughs.

\subsection{Participant Recruitment}
\label{Recruitment}
\label{user-study1}
We recruited 35 undergraduate students and recent college graduates who did not study nor intend to study any of the 5 science fields. We chose non-experts because they represent everyday people, who are our target audience for science communication on social media. Because the Tweetorials that we aim to emulate are US-centric in the examples and language used, we disqualified participants who do not self-identify as culturally American and whose first language is not English. The participants are from 12 universities and liberal arts colleges in the United States, with an average age of 22, and a gender distribution of 4 males, 29 females, and 2 non-binary individuals. We distributed the recruitment survey via school mailing lists, Slack workspaces, Discord channels, and snowball sampling \cite{goodman_snowball_1961} among schoolmates of the participants. Each participant was compensated \$27 dollars. The study was approved by our institutional IRB. 

\subsection{Survey Procedure}
Following recruitment, each qualified participant attended an onboarding session with an experimenter on Zoom. The experimenter explained the study procedure and data collection, and acquired participants' consent. The experimenter shared a sample Twitter thread and survey for the participant to answer and explained the Likert-scale rating criteria. The participants thought aloud while answering the sample survey, justified their rating decisions, and asked clarifying questions before completing the survey on their own. Participants' justifications for their ratings on the sample science narratives aligned with the defined definitions. 

We implemented our survey on Qualtrics, an online survey platform. On the first page, participants are asked to read a Twitter thread. The page included a one-minute timer to prevent participants from skimming or skipping the reading process. After reading the Twitter thread, the participant advances to the next page and completes four Likert-scale questions that correspond to the four evaluation dimensions (engaging, relatable, understandable, and easy-to-follow, see Section \ref{survey_design}), on the scale from 1 to 5 (1-Strongly Disagree, 5-Strongly Agree). This process repeats 15 times for the 15 science explanations for each participant. 

To minimize bias from prior knowledge, we employed a between-subjects design, ensuring that each participant only read each topic once, and no participant read the same topic under two different conditions. Each participant read a randomized selection of Twitter threads across all conditions and topics. Each Twitter thread was evaluated by 7 participants to achieve a statistical distribution and reduce individual bias.

\subsection{Data Collection and Analysis}
\label{Data}

\subsubsection{Survey Design}
\label{survey_design}
Each science explanation was evaluated on 4 different questions: (1) I find the thread engaging, (2) I find the thread relatable, (3) I find the thread understandable, and (4) I find the thread easy to follow. \textit{Engaging} refers to the language, while \textit{relatable} refers to the example used. \textit{Understandable} refers to the presence of technical jargon, while \textit{easy-to-follow} refers to the sequence and flow of the narrative. The onboarding session demonstrated that participants were able to differentiate between each of the dimensions. We sum the participant's ratings for each Likert-scale question to create an overall score for the given science explanation (20-point max score).

\subsubsection{Data Analysis}
\textbf{Quantitative Analysis: Survey Data} 

Our main goal was to test whether the inclusion of examples, walkthroughs, and personal language influenced readers' engagement and understanding. We used the overall score (sum of the 4 Likert responses) for the given science explanation in our analysis. Because each participant did not read the same topic twice and sampled across multiple STEM fields, our data exhibited a hierarchical structure: each rating is linked to a specific participant and a specific topic.

We used a Generalized Linear Mixed Model (GLMM) for our analyses. The GLMM framework allowed us to isolate how each technique affected the ratings for each science explanation based on the overall score of the Likert ratings, while statistically controlling for other factors. We can estimate the direction and magnitude of the effectiveness of each technique (examples, walkthroughs, and personal language) relative to explanations without them. We also accounted for individual differences in baseline rating tendencies by including random intercepts for each participant because people may vary in their strictness or leniency when rating explanations. We also included random intercepts for each of the five STEM topics to account for potential differences across fields (e.g., participants might rate Computer Science explanations differently than Physics explanations). The GLMM model allowed us to compare pairs of explanations—those with a given technique versus those without—to determine whether the presence of each technique had a significant effect on the ratings.

\textbf{Qualitative Analysis: Followup Interviews}
To gain more nuanced insights into participant preferences in the survey data, we randomly reached out to 17 of the 35 participants for follow-up interviews. 8 of 17 participants opted-in for a 15-minute follow-up interview on Zoom. The participants are compensated \$10 total. In the follow-up interview, we asked the participants to re-read a science explanation they had previously read and explain in detail specific instances from the science narrative that influenced their Likert ratings.  The interviews were recorded and transcribed before three researchers conducted a thematic analysis. We used grounded theory to derive insights from the interview transcripts \cite{charmaz_constructing_2006}. The first author read through all interview transcripts, inductively derived a preliminary set of codes, and then grouped the codes based on themes. The first author and two additional authors collaboratively reviewed and refined the themes until a consensus was reached.

\section{Results: Reader Study}

Overall, we collected 105 ratings on 75 different science explanations. This consisted of 35 different annotators who evaluated each science explanation on 4 different Likert-scale questions. We conduct a quantitative analysis of the survey data followed by a qualitative analysis of interview transcripts (n=8) for individual reader's preferences for science communication to contextualize nuances to the quantitative data.

\subsection{Usage of Example Preferences}
To evaluate H1:Example, we compare experimental condition EWP to baseline condition WP. We found that participants prefer reading explanations with an example (EWP) over without an example (WP).
The results presented in Table \ref{tab:glmm_results_H1} illustrate the effects of the two conditions on the total score of the four survey questions (max 20 points) within a Generalized Linear Mixed Model (GLMM) framework. The experimental condition, EWP, received a score of 15.831 which shows a statistically significant increase in user preference when compared to the baseline condition of 13.003 (p < 0.0001). The effect size is 2.828 which shows that readers rated narratives with an example (EWP) almost 3 points higher than narratives without an example (WP). Additionally, the random effects analysis reveals a large group variance of 2.626 suggesting the differences among individual participants play a significant role and a small fields variance at 0.062, the specific field of study has a minimal impact. 

From our qualitative interviews with readers, we found that overall, readers reported that having an example was helpful. 4 of 8 readers stated the example helped them understand the importance of a topic (P2, P4, P7, P8). 5 of 8 found that examples helped them reflect and understand their own experiences better (P1, P2, P3, P4, and P7). However, not all readers needed an example to help them understand the topic. 3 of 8 participants mentioned how the example felt unnecessary and detracted from the content of the explanation (P1, P2, P8). When reading about depth-first search, P1 mentioned how the given example of navigating through the Botanical Garden felt unnecessary: \textit{``The concept in and of itself is interesting. I would have just read about that. I don't really need any more context.''} For P8, when learning about thin film interface through the example of a child playing with bubbles, he remarked that \textit{``[the example] doesn't really pertain to me in any way,''} demonstrating how certain examples might not resonate with particular audiences.

These results demonstrate that while the use of examples generally help in providing additional context and grounding for a science narrative, the effectiveness of an example depends on the science topic and the readers' personal preferences. \textbf{Scientists do not need to avoid examples out of fear of distracting or alienating readers. Instead, the use of examples in science narratives is situational and scientists should have the agency to compare how different narratives with and without an example might resonate with readers depending on the topic and audience.}

\begin{table}[h]
    \centering
    \begin{tabular}{@{}lccccc@{}}
        \toprule
        \textbf{Effect}                  & \textbf{Score (max. 20)}& \textbf{Coefficient} & \textbf{Standard Error} & \textbf{z-value} & \textbf{p-value} \\ \midrule
        Intercept[WP]                        & 13.003 & 13.003               & 0.483                   & 26.937            &           \\ \midrule
        CONDITION[EWP]       & 15.831 & 2.828                & 0.546                   & 5.183             & \textbf{0.000}          \\ \midrule
        \textbf{Random Effects}   &        &                       &                          &                   &                   \\ 
        Group Variance    &        & 2.626                & 0.752                   &                   &                   \\ 
        Fields Variance      &            & 0.062                & 0.543                   &                   &                   \\ \bottomrule
    \end{tabular}
    \caption{GLMM Results for H1:Examples show that readers demonstrate a \textbf{statistically significant} preference for science narratives \textbf{with an example} over narratives without an example.}
    \label{tab:glmm_results_H1}
\end{table}

\subsection{Step-by-Step Walkthrough Preferences}
To evaluate H2:Walkthrough, we compare experimental condition EWP-NoFewShot to baseline condition EP-NoFewShot. We found that readers had a slight preference for explanations with a walkthrough compared to without a walkthrough, with a considerable variance on the fields of study.

Table \ref{tab:glmm_results_H2} summarizes the findings from a GLMM analysis. The experimental condition, EWP-NoFewShot received a score of 15.965 which was not statistically significant compared to 14.729, the baseline score for EP-NoFewShot (p = 0.057). As such, there is no significant difference in how readers rated explanations with and without walkthroughs. The random effects analysis shows a minimal group variance of 0.002, indicating little variability among participants. However, the field variance is considerable at 2.566, suggesting that participants prefer having a walkthrough for some STEM topics but prefer having no walkthrough for other STEM topics. 

From our qualitative findings, we find that the use of a walkthrough helped helped establish a structure for the explanation (P2, P3, P8), provide evenly paced information (P1, P3, P6), and helped them follow through with the science (P4, P6, P7). But for 3 readers, the use of a walkthrough sometimes made the explanation feel overly explanatory or repetitive (P2, P3, P7). For the topic of Walker's Action Decrement Theory, P2 said that the walkthrough narrative was \textit{``just overly explanatory for a concept that is very intuitive,''} demonstrating how for certain topics a walkthrough might not be necessary. 4 participants (P2, P4, P5, P6) preferred science explanations without a walkthrough because they provided multiple different perspectives to understand the topic. P4 stated that the explanation without a walkthrough broke down the topic of the curtain wall system into smaller, self-contained chunks of information that made reading the explanation \textit{``less intimidating.''}

These results demonstrate that while a step-by-step walkthrough was helpful in explaining the science, but for certain STEM topics the walkthrough might not be necessary. \textbf{Scientists should use their own best judgment in evaluating when added detail supports clarity and when a more streamlined explanation may be sufficient for the topic at hand.}

\begin{table}[h]
    \centering
    \begin{tabular}{@{}lccccc@{}}
        \toprule
        \textbf{Effect}                 & \textbf{Score (max. 20)} & \textbf{Coefficient} & \textbf{Standard Error} & \textbf{z-value} & \textbf{p-value} \\ \midrule
        Intercept[EP-NoFewShot]               & 14.729         & 14.729               & 0.439                   & 33.537            &        \\ \midrule
        CONDITION[EWP-NoFewShot]    & 15.695    & 0.966                & 0.508                   & 1.902             & 0.057          \\ \midrule
        \textbf{Random Effects}   &       &                       &                          &                   &                   \\ 
        Group Variance     &       & 0.002                & 0.618                   &                   &                   \\ 
        Fields Variance        &          & 2.566                & 0.506                   &                   &                   \\ \bottomrule
    \end{tabular}
    \caption{GLMM Results for H2:Walkthrough shows that there is no significant difference between readers' preferences for science narratives with or without walkthroughs.}
    \label{tab:glmm_results_H2}
\end{table}

\subsection{Personal Language Preferences}
To evaluate H3:Personal Language, we compare experimental condition EWP to baseline condition EW. We found that participants prefer reading explanations with personal language over without personal language. 

Table \ref{tab:glmm_results_H3} summarizes the findings from a GLMM analysis. The experimental condition, EWP, received a statistically significantly higher score of 15.883 compared to 13.973 for the baseline condition, EW (p < 0.0001). The effect size is 1.910 which means that readers rate explanations with personal language almost 2 points higher than explanations without personal language. Additionally, there is a group variance of 1.392 indicating a moderate level of variability among participants and a fields variance of 1.361 showing a moderate variance among STEM topics. 

From the qualitative findings, we see that science narratives with personal language helped establish a writer/reader connection (P1, P3, P5) and helped make science topics more approachable (P2, P3, P6, P7). But not all readers preferred narratives with personal language. 2 readers preferred reading science explanations with no personal language because they believed personal language was unnecessary (P1, P8): \textit{``[the explanation is] full of fluff coming from a personal perspective that I didn’t really care about''} (P1). P8 mentioned that he did not need personal language for engagement when reading about science, though he hypothesized that other readers \textit{``[might] need the personal aspects to get them to read something that they wouldn't otherwise be interested in.''} Depending on a reader's inclination towards science, personal language may or may not be necessary to help them engage with these science narratives. Even for readers who generally do prefer science narratives with personal language, there's a balance between narratives with just enough and too much personal language (P2, P7). P2 compared two narratives with personal language saying how for one example \textit{``the language was not distracting'' } while the other example \textit{``had a ton of hashtags and full of like, fluff.''}

While personal language is helpful in making science more engaging, the sentiment is not shared by all readers. Even for readers who generally enjoy personal language in science narratives, there's a balance of how much personal language should be incorporated before it detracts from the science. \textbf{These findings highlight that preferences are not binary; rather, they fall along a spectrum, where different readers find varying degrees of personal language to be effective for a science narrative.}

\begin{table}[h]
    \centering
    \begin{tabular}{@{}lccccc@{}}
        \toprule
        \textbf{Effect}        & \textbf{Score(max. 20)}          & \textbf{Coefficient} & \textbf{Standard Error} & \textbf{z-value} & \textbf{p-value} \\ \midrule
        Intercept[EW]      & 13.973                  & 13.973               & 0.463                   & 30.175            &           \\ \midrule
        CONDITION[EWP]   &15.883     & 1.910                & 0.526                   & 3.632             &\textbf{ 0.000   }       \\ \midrule
        \textbf{Random Effects}   &       &                       &                          &                   &                   \\ 
        Group Variance     &       & 1.392                & 0.717                   &                   &                   \\ 
        Fields Variance       &           & 1.361                & 0.548                   &                   &                   \\ \bottomrule
    \end{tabular}
    \caption{GLMM Results for H3:Personal Language show that there is a \textbf{statistically significant} difference in readers preference for science narratives that \textbf{contain personal language} over narratives without personal language.}
    \label{tab:glmm_results_H3}
\end{table}

\section{Writer Study: Methodology}
From the reader study, we found that readers had personal preferences for how different public science communication strategies (example, walkthrough, and personal language) were used. Readers said that examples and walkthroughs helped provide more a scaffolded understanding of the STEM topics, but not all topics or readers needed the same level of elaboration. For personal language, while many readers preferred it, there was a spectrum of preferences regarding how much personal language was needed in science narratives. 

The findings from the reader study illustrate that science communication strategies cannot be applied as one-size-fits-all approach: use or not use examples/walkthroughs/personal language. Instead, the effectiveness of these strategies depends on both the STEM topic and the individual reader’s preferences. Given the diverse and contextual preferences that readers had for these science communication strategies, static writing guidelines do not accommodate these preferences. As such, we designed a public science writing support tool that assists scientists in negotiating how to adopt science communication strategies, with the goal of helping scientists make more informed and intentional choices about how to communicate STEM topics. 

In designing a system for writers, we use a two step process that helps writers first \textbf{structure}, then \textbf{style} their science narrative. Structure pertains to how the science is communicated through examples and walkthroughs. Style refers to the tone and voice of the narrative. Our system uses contrastive examples to help writers navigate structure (one, none, and many examples) and style (with and without personal language) options. We conduct a user study with 10 scientists to understand how seeing different examples might help writers by providing them with contrastive versions of narratives that vary in structure (e.g., examples and walkthroughs) and style (e.g., personal language).

Specifically we are interested in the following research questions:

\textbf{RQ1:} How does seeing structure options (\textit{One Example}, \textit{No Example}, \textit{Many Examples}) help writers consider different framing strategies for writing science for social media?

\textbf{RQ2:} How does seeing style options (\textit{With Personal Language}, \textit{Without Personal Language}) help writers consider different communication styles when writing science for social media?

\subsection{Participants}
We recruited 10 PhD-level researchers interested in communicating their research to the public on social media. Participants are from 2 universities, an average age of 25, with a gender distribution of 9 males and 1 female (Table \ref{tab:my-table}). We advertised the study to students in research labs through school mailing lists, Slack workspaces, and snowball sampling among lab mates of the participants. Their expertise spans various CS research areas, including natural language processing, programming languages, and social computing. The study was conducted over Zoom and took approximately 2 hours. Each participant is compensated \$40 dollars total. The study was approved by our institutional IRB. 

\begin{table}[]
\begin{tabular}{|c|c|c|}
\hline
\textbf{ID} &  \textbf{Field of Expertise}              & \textbf{Research Experience (years)} \\ \hline
1                    & Computer Science and Journalism          & 2                            \\ \hline
2                    & Artificial Intelligence and Neuroscience & 2                            \\ \hline
3                    & Human-Computer Interaction               & 2.5                          \\ \hline
4                  & Natural Language Processing              & 2                            \\ \hline
5                    & Natural Language Processing              & 3                            \\ \hline
6                       & Programming Languages                    & 5                            \\ \hline
7                      & Computer Security                        & 2.5                          \\ \hline
8                      & Quantum Computing                        & 3                            \\ \hline
9                  & Human-Computer Interaction               & 3                            \\ \hline
10                   & Computer Science Education               & 7                            \\ \hline
\end{tabular}
\caption{Participant Demographics for Writer Study}
\label{tab:my-table}
\end{table}

\subsection{Writing Study Procedure and Analysis}

\subsubsection{Study Procedure}
Each session includes a pre-study presentation, an interface demonstration, and two writing sessions with a 15-minute break in between. The experimenter used a short presentation to educate participants about science communication on social media, a background on Tweetorials, and different techniques and examples for science communication on social media. Next, the experimenter demonstrated the study web interface and explained how it worked. 

% \subsubsection{Writing Study}
When the participant was ready for the writing session, the experimenter shared with the participant a URL link to the study web interface. The experimenter started screen and audio recording upon the participant's verbal consent. Participants use the study interface once for each writing session: first to write about the predetermined topic of merge sort and then to write about a topic of their choice. The experimenter implemented think-aloud protocol and encouraged the participant to voice their thought process, reasoning behind their structure and style choices, and editing decisions. After each writing session, the experimenter conducted a semi-structured interview with the participant to understand how seeing different structures and styles influenced their writing choices. Some sample questions are: 
How did you arrive at your choice of structure/style?
How did seeing the options change your preference of structure/style?
How have reading the unselected options help you decide which direction to write in?

\subsection{Study Interface}
\label{sec:writer_workflow}

\begin{figure}
    \centering
    \includegraphics[width=1.0\linewidth]{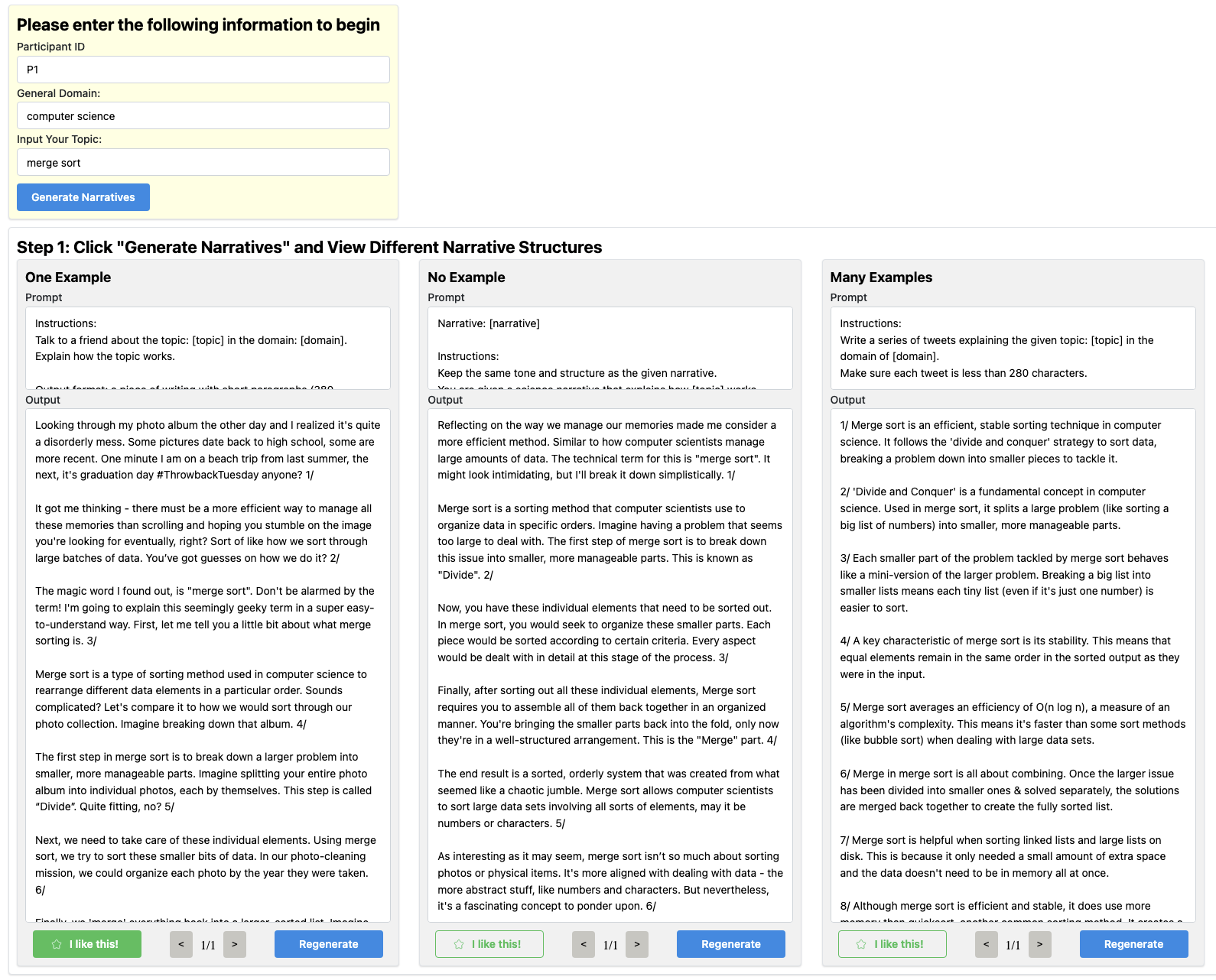}
    \caption{Study interface: Viewing different structure options (1st of 3 steps)}
    \label{fig:step1}
\end{figure}

\begin{figure}
    \centering
    \includegraphics[width=1.0\linewidth]{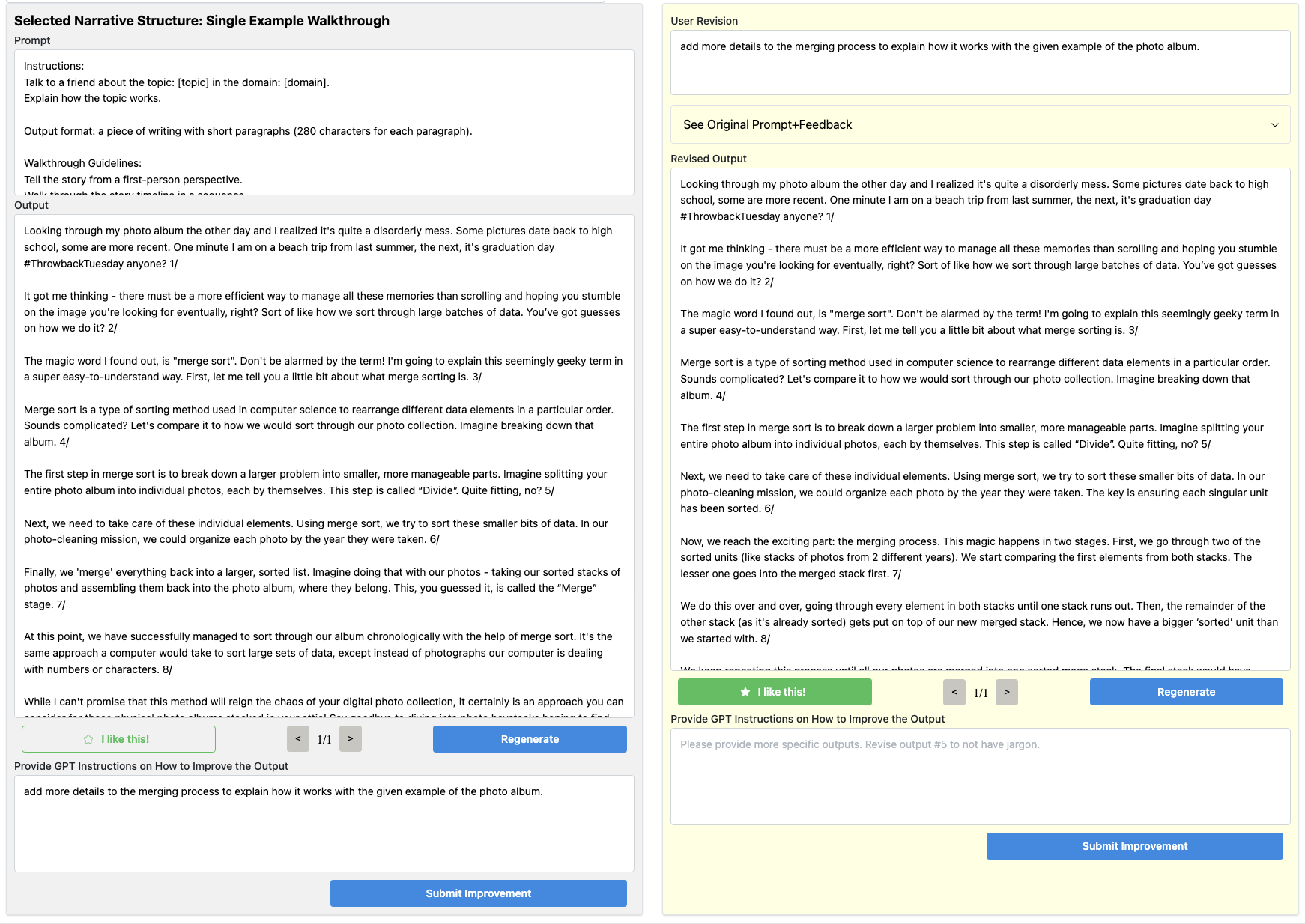}
    \caption{Study interface: Editing narrative structure (2nd of 3 steps)}
    \label{fig:step2}
\end{figure}

\begin{figure}
    \centering
    \includegraphics[width=1.0\linewidth]{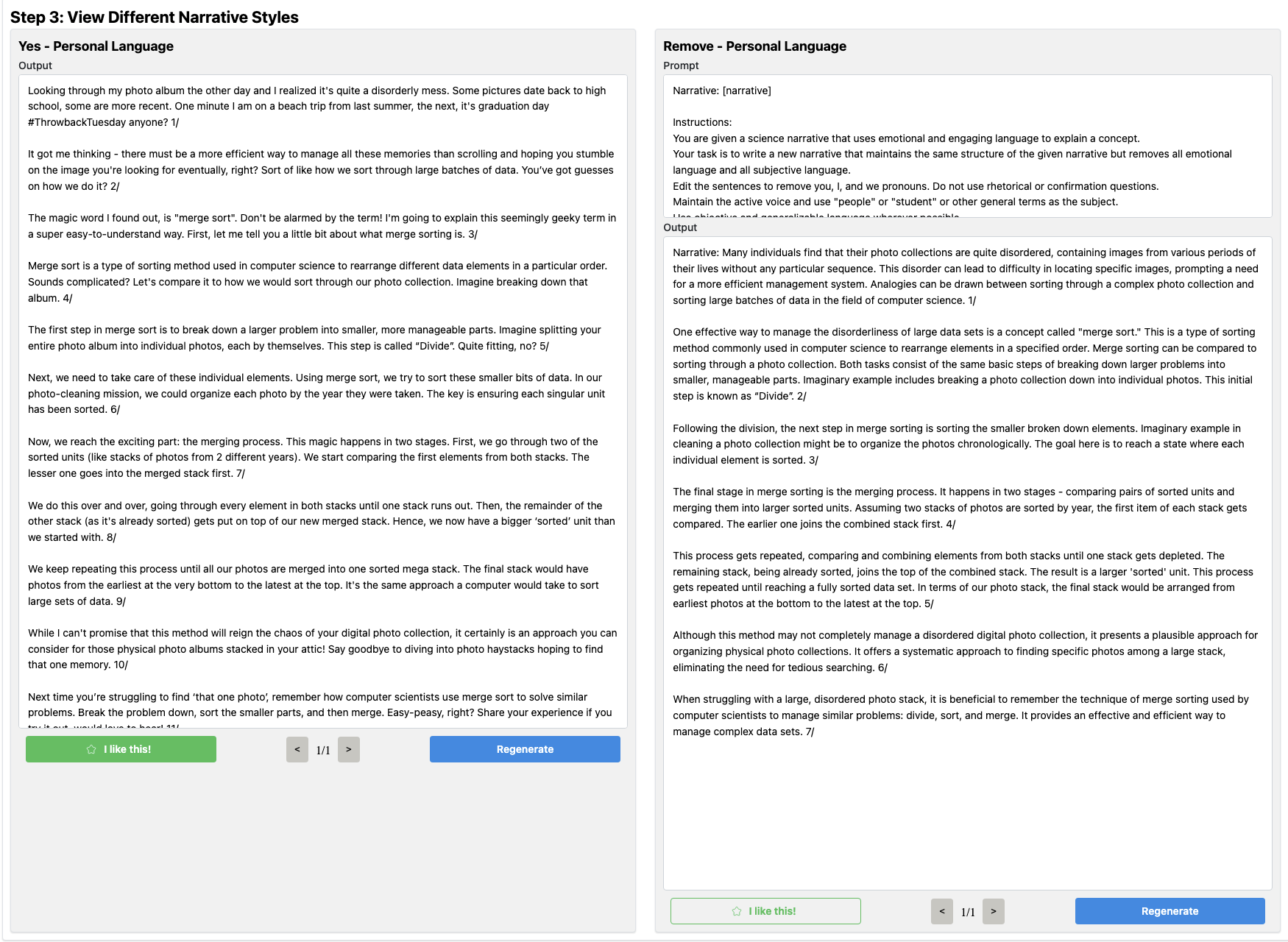}
    \caption{Study interface: Viewing different style options (3rd of 3 steps)}
    \label{fig:step3}
\end{figure}

We built a web interface that guides writers through a workflow with LLM generations and editing functionalities (Figure \ref{fig:step1}, Figure \ref{fig:step2}, Figure \ref{fig:step3}). The workflow includes the following 4 steps:

\textbf{Step 1: Structure Options} Users first enter their domain and topic to generate the 3 different structure options. After the generation, 3 columns are displayed side-by-side with the corresponding LLM prompts: \textit{One Example}, \textit{No Example}, and \textit{Many Examples} (Figure \ref{fig:step1}. The user chooses one of the three to proceed with. Writers can merge certain paragraphs from two structure options by copying and pasting between the columns before proceeding. 

\textbf{Step 2: Selected Structure Feedback and Edits} The second part of the interface (Figure \ref{fig:step2}) allows the writer to iterate on the selected structure by providing feedback instructions to an LLM or making manual edits directly in the textbox (copy, paste, delete, and type). The writer is asked to focus only on editing structural aspects of the text such as technical accuracy, content, and sequence. When they are satisfied with the structure, they proceed to the next step.

\textbf{Step 3: Language Style Options} The third part of the interface compares different style options, it has 2 columns displayed side by side (Figure \ref{fig:step3}). On the left is the writer's selected and edited draft from the previous step which contains personal language and the right is the draft without personal language. Writers select which narrative they want to proceed with to the next step. Writers can merge certain paragraphs from both options by copying and pasting between the columns before proceeding.

\textbf{Step 4: Final Edits}
The last part of the interface displays the draft from Step 3 (not shown). The writer refines and finalizes the writing until they are satisfied and ready to share it on social media.

% \subsubsection{Semi-Structured Interview}
\subsubsection{Data Analysis}
We analyze interview transcripts and video recordings to understand quantitative and qualitative aspects of participants' choices and reasoning when presented with different options. We analyzed participants' choice of structure (\textit{One Example}, \textit{No Example}, \textit{Many Examples}) and style options (\textit{With Personal Language} and \textit{Without Personal Language}), participants' editing actions and LLM feedback prompts, iterations of writing generations and refinements, as well as the final writings. 

One researcher independently conducted a bottom-up, open-coding approach to data analysis \cite{charmaz_constructing_2006}. Then, the researcher worked with two other researchers to iterate on the codes, discuss their similarities and differences as part of a comparative analysis \cite{merriam2015qualitative}, and leveraged them in an affinity diagramming process \cite{holtzblatt2017affinity}. The researchers determined that they reached code saturation when no researcher could identify new codes or arrive at new interpretations.

\section{Results: Writer Study}

\subsection{\textbf{RQ1}: Different Narrative Structures Helps Scientists Identify What Content to Include (or Not Include)}

Scientists chose different narrative structures to work off based on the STEM topic they were writing about, their individual preferences, or their judgments about what would be most accessible to an everyday audience. Most science narratives used the \textit{One Example} narrative structure to iterate on (13 of 20). The \textit{Many Examples} strategy was chosen 5 of 20 times. The least used structure was \textit{No Examples} (2 of 20).  Choices made by participants and topic are found in Table \ref{tab:writer-structure-preferences}. From the interviews, scientists' choice of narrative structure depended on the topic and the content that they wanted to include in their science narrative. Scientists chose the  \textit{One Example} option because the structure provided a clear sequence of explanation for scientists follow. Other scientists chose to merge between the \textit{One Example} and \textit{Many Examples} structure to provide additional examples to explain certain aspects of a STEM topic. One writer chose to use the \textit{No Example} option because it provided a strong technical explanation for scientists to build off of (P3).

Scientists' preferences and beliefs about how to structure science narratives were revealed or challenged when they saw contrasting structure options (P2, P3, P5, P7, P8, P9, P10). Seeing the comparison of different narrative structures, helped P9 solidify his initial feeling that \textit{``one example is a good way to engage people.''} In contrast, P7's initial inclination to use the \textit{One Example} narrative structure was challenged after seeing the different structure options. P7 chose to use the \textit{Many Examples} narrative structure because he was \textit{``able to see what other things I might actually want to include when I'm trying to explain it, and what other possibilities there are for explaining.''} Seeing multiple different narrative structures prompted P6 to more carefully evaluate which structure option was most suitable for his topic of formal verification:\textit{``I'm not sure if I will pick the one example narrative, because it's a hard topic to explain to a general audience. I'll look through the other options more to see if there's something more suitable for this topic.''} Contrasting narrative structures helped confirm existing writer preferences and prompted some scientists to reconsider which narrative structure was most suitable for their topic.

The narrative structures that scientists did not utilize often provided a negative example that highlighted what to avoid when explaining science to an everyday audience for 8 scientists (P1, P2, P3, P5, P6, P7, P8, P9). P6 said \textit{Many Examples} narrative had a `\textit{`rigorous depiction of what the software is, [which can be hard] for a general audience to imagine what that means."} For P9, seeing the \textit{No Example} narrative helped him think from the perspective of a reader and consider \textit{``why [a reader] would decide to click on the [No Example narrative]. Unless the reader was looking to understand what merge sort was, the [No Example narrative] would not help the everyday reader get into the topic.''} Even though scientists may not have chosen a particular narrative structure to use, the act of reading through different narrative options helped scientists clarify how to present their science topic for an everyday audience.

Four scientists chose to merge two narrative structures together to combine the different strengths of each structure (P2, P4, P6, P10). For the topic of multiplicative weights update, P10 merged elements from \textit{Many Examples} into a base narrative with \textit{One Example} to emphasize the dimensions of online learning and regret minimization which are \textit{``important facts about the topic, but wasn't brought up in the [\textit{One Example}] option.''} Because the \textit{Many Examples} option provided him with different angles to motivate or contextualize the topic, the selected paragraphs from \textit{Many Examples} could be \textit{``slotted in at the end."} Seeing different narrative structures helped scientists balance trade-offs such as technical content and clarity.

\textbf{Seeing different structure examples helped scientists identify the science content that they might want to include and provide a scaffold to easily integrate different content that they should include for the STEM topic they were writing about.}

\begin{table}[H]
\resizebox{\textwidth}{!}{%
\begin{tabular}{|c|c|c|c|c|c|c|}
\hline
{\textbf{Participant ID}} & {\textbf{Domain}} & {\textbf{Topic}} & {\textbf{One Example}} & {\textbf{No Example}} & {\textbf{Many Examples}} & \textbf{Merged}\\ \hline\hline

\multirow{2}{*}{P1} & {Computer Science} & {Merge Sort} &  &  & {\checkmark} & \\ \cline{2-7}
& { Statistics} & { Gradient Linear Additive Models} & {\checkmark} &  &  & \\ \hline

\multirow{2}{*}{P2} & {Computer Science} & {Merge Sort} & & & & {\begin{tabular}[c]{@{}c@{}}Primary: One Example\\ (Secondary: Many Examples)\end{tabular}}                          \\ \cline{2-7}
& {Computer Science} & {Gradient Descent} & {\checkmark} & & & \\ \hline
 
\multirow{2}{*}{P3} & {Computer Science} & {Merge Sort} & & {\checkmark} & &  \\ \cline{2-7}
& {Qualitative Analysis} & {Ordinal Regression} & & {\checkmark} & & \\ \hline
 
\multirow{2}{*}{\vspace{-0.21in} P4} & {Computer Science} & {Merge Sort} & & & & {\begin{tabular}[c]{@{}c@{}}Primary: One Example\\ (Secondary: Many Examples)\end{tabular}}  \\ \cline{2-7}
& {Natural Language Processing} & {Controlled Decoding} & & & & {\begin{tabular}[c]{@{}c@{}}Primary: Many Examples\\ (Secondary: One Example)\end{tabular}} \\ \hline

\multirow{2}{*}{P5} & {Computer Science} & {Merge Sort} & {\checkmark} & & & \\ \cline{2-7}
& {Natural Language Processing} & {Word Embeddings} & & & {\checkmark} & \\ \hline
 
\multirow{2}{*}{\vspace{-0.2in}P6} & {Computer Science} & {Merge Sort} & {\checkmark} & & & \\ \cline{2-7}
& {Programming Languages} & {Formal Verification} & & & & \makecell{Primary: One Example \\ (Secondary: Many Examples)} \\ \hline

\multirow{2}{*}{P7} & {Computer Science} & {Merge Sort} & {\checkmark} & & &  \\ \cline{2-7}
& {Natural Language Processing} & {Embedding Space} & & & {\checkmark} & \\ \hline
 
\multirow{2}{*}{P8} & {Computer Science} & {Merge Sort} & {\checkmark} & & & \\ \cline{2-7}
& {Quantum Algorithms} & {Grover's Algorithm} & & & {\checkmark} & \\ \hline

\multirow{2}{*}{P9} & {Computer Science} & {Merge Sort} & {\checkmark} & & & \\ \cline{2-7}
& {Tangible User Interfaces} & {4D Printing} & {\checkmark} & & & \\ \hline
 
\multirow{2}{*}{\vspace{-0.18in} P10} & {Computer Science} & {Merge Sort} & {\checkmark} & & & \\ \cline{2-7}
& {Optimization Algorithms}     & {Multiplicative Weights Update} & & & & {\begin{tabular}[c]{@{}c@{}}Primary: One Example\\ (Secondary: Many Examples)\end{tabular}}  \\ \hline\hline

& & \textbf{Total Count:} & \textbf{\begin{tabular}[c]{@{}c@{}}9\\ One Example\end{tabular}} & \textbf{\begin{tabular}[c]{@{}c@{}}2 \\ No Example\end{tabular}} & \textbf{\begin{tabular}[c]{@{}c@{}}4\\ Many Examples\end{tabular}} & \textbf{\begin{tabular}[c]{@{}c@{}}5\\ Merged\end{tabular}} \\\hline            
\end{tabular}%
}
\caption{Participants, the topic they wrote on, and the corresponding narrative structure that they chose. When merging two narratives, ``primary'' denotes the narrative structure that the participant used as the base and ``secondary'' means that the writer incorporated elements from this narrative into the primary narrative.}
\label{tab:writer-structure-preferences}
\end{table}

\subsection{\textbf{RQ2}: Different Style Options Help Scientists Balance Personal Preferences with Social Media Communication}

The most common narrative strategy that scientists used was \textit{With Personal Language} (12 of 20). Merging elements from the \textit{With Personal Language} and \textit{Without Personal Language} style was the second most common action that scientists performed (6 of 20). Only 2 of 20 narratives used narratives without personal language. From interviews, scientists generally preferred using science narratives with personal language and editing the LLM-generated language to match their own voice. Many scientists found that comparing between narratives with and without personal language helped them identify and adopt useful techniques they might not have otherwise used in their own writing. Even though most scientists choose to iterate on science narrative with personal language, some scientists found that working off of narratives without personal language helped them clarify the science before adding back personal language.

Scientists' recognized the importance of personal language and integrated personal language into their science narrative after seeing narratives with and without personal language. Even when scientists were initially hesitant about using personal language, seeing narratives with and without personal language helped four scientists adopt personal language in their final narratives (P1, P3, P9, P10). For P7, seeing GPT-generated personal language helped him adapt to the genre of science communication on social media: \textit{``I think the enthusiasm, while it may not necessarily be the way that I would have written this, feels like it's a more engaging way of trying to explain this to people on the internet.''}  P1 was initially resistant to using personal language, but after reading the ``Without Personal Language'' narrative, P1 said \textit{``the comparison between with and without personal language, was good as a recall to me that this type of personal language is important in a twitter threads.''} Contrasting examples with and without personal language helped scientists recognize and adopt personal language, even when they were initially hesitant to including it.

Furthermore, seeing the two contrasting style options helped 7 scientists balance technical content with personal exposition (P1, P2, P3, P5, P7, P8, P9). Narratives with personal language often contained more personal exposition about specific details or setting the scene, so seeing a narrative without personal language helped scientists \textit{``introduce the premise a lot more concisely''} (P7). P9 said that \textit{``because the narrative without personal language is more technical as I'm editing the narrative with personal language, I'll go over to the right [without personal language] if I feel like more of an explanation is needed, and see if there's a technical detail that I can bring in.''} The tradeoff between personal language and concision was highlighted through the side-by-side comparison of science narratives with and without personal language. 

\textbf{Seeing contrasting examples of science narratives with and without personal language helped scientists (1) adopt the use of personal language or (2) strike a balance between technical concision and descriptive problem statements.}

\begin{table}[H]
\resizebox{\textwidth}{!}{%
\begin{tabular}{|c|c|c|c|c|c|}
\hline
{\textbf{Participant ID}} & {\textbf{Domain}} & {\textbf{Topic}} & {\textbf{With Personal Language}} & {\textbf{Without Personal Language}} & {\textbf{Merged Narrative Styles}} \\ \hline\hline
  
\multirow{2}{*}{P1} & {Computer Science} & {Merge Sort} & {\checkmark} & & \\ \cline{2-6}
& {Statistics} & {Gradient Linear Additive Models} & & & Primary: Without Personal Language \\ \hline 

\multirow{2}{*}{P2} & {Computer Science} & {Merge Sort} & & {\checkmark} & \\ \cline{2-6}
& {Computer Science} & {Gradient Descent} & & & Primary: With Personal Language \\ \hline 
 
\multirow{2}{*}{P3} & {Computer Science} & {Merge Sort} & & & Primary: With Personal Language \\ \cline{2-6}
& {Qualitative Analysis} & {Ordinal Regression} & & & Primary: Without Personal Language \\ \hline 
 
\multirow{2}{*}{P4} & {Computer Science} & {Merge Sort} & {\checkmark} & & \\ \cline{2-6}
& { Natural Language Processing} & {Controlled Decoding} & {\checkmark} & & \\ \hline 

\multirow{2}{*}{P5} & {Computer Science} & {Merge Sort} & {\checkmark} & &  \\ \cline{2-6}
& {Natural Language Processing} & {Word Embeddings} & & & Primary: With Personal Language \\ \hline 
 
\multirow{2}{*}{P6} & {Computer Science} & {Merge Sort} & {\checkmark} & & \\ \cline{2-6}
& {Programming Languages} & { Formal Verification} & {\checkmark} & & \\ \hline 

\multirow{2}{*}{P7} & {Computer Science} & {Merge Sort} & & &  Primary: With Personal Language \\ \cline{2-6}
& {Natural Language Processing} & {Embedding Space} & {\checkmark} & &  \\ \hline 
 
\multirow{2}{*}{P8} & {Computer Science} & {Merge Sort} & {\checkmark} & & \\ \cline{2-6}
& { Quantum Algorithms} & { Grover's Algorithm} & & {\checkmark} & \\ \hline 

\multirow{2}{*}{P9} & {Computer Science} & {Merge Sort} & {\checkmark} & &  \\ \cline{2-6}
& {Tangible User Interfaces} & {4D Printing} & {\checkmark} & &  \\ \hline 
 
\multirow{2}{*}{P10} & {Computer Science} & {Merge Sort} & {\checkmark} & & \\ \cline{2-6}
& {Optimization Algorithms} & {Multiplicative Weights Update} & {\checkmark} & & \\ \hline \hline

& & \textbf{Total Count:} & \textbf{\begin{tabular}[c]{@{}c@{}}12\\ With Personal Language\end{tabular}} & \textbf{\begin{tabular}[c]{@{}c@{}}2 \\ Without Personal Language\end{tabular}} &  \textbf{\begin{tabular}[c]{@{}c@{}}6\\ Merged\end{tabular}} \\ \hline
\end{tabular}%
}
\caption{Participants, the topic they wrote on, and the corresponding narrative style that they chose. For merged narratives, ``primary" denotes the narrative style that the participant chose as their base narrative, and scientists incorporated elements from the remaining style option into the primary narrative.}
\label{tab:writer-preferences-style}
\end{table}

\section{Discussion}

\subsection{Narrative Structure Options Help Scientists Identify Science Content to Include}
Our findings demonstrate that exposing scientists to multiple narrative structures plays a role in shaping what content scientists choose to include or exclude. Since not all STEM topics require the same level of detail, providing multiple different structure options allows scientists to identify the level of specificity that is important for their STEM topic. Seeing different narrative structure examples helps scientists identify gaps in the explanation to add more detail---by including alternative examples to explain the topic---or to remove extraneous details to support readers in understanding the science. Furthermore, seeing different narrative structures puts the writer into the perspective of a reader and helps them consider what types of science content would be the most beneficial to an everyday audience. For scientists who have already amassed technical expertise in their domain, using AI to present initial drafts allows them to focus on evaluating the trade-offs between different structure options such as technical detail and clarity. 

This finding relates to a broader usability heuristic of ``recognize over recall'' which emphasizes the use of visual cues to help with memory retrieval as opposed to remembering something from scratch \cite{gigerenzer2011recognition, nielsen10usability}. This design heuristic is important because it reduces the amount of cognitive effort required by the user to perform the task. In the context of science communication, presenting scientists with explicit structural alternatives acts as a recognition aid: instead of having to recall different ways of structuring an explanation, scientists can recognize, compare, and adapt pre-existing narrative structure options. This not only lowers the effort required for scientists to use effective communication practices but also helps scientists to critically evaluate the tradeoffs between different structure options to find one (or a combination of structures) to best communicate their topic to an everyday audience.

\subsection{Presenting Personal Language as a Continuum and not a Binary}
Prior research has shown that telling scientists to use personal language is not enough to motivate them to adopt this science communication strategy \cite{gero2021makes}. We found that presenting starkly contrasting narratives with and without personal language helped scientists identify how much personal language they wanted to include, even though it was typically in between the examples. Even scientists who were initially hesitant to use personal language in their science narrative adopted some personal language in their final science narrative because they recognized how narratives without personal language were overly technical and not engaging. Contrasting examples allowed scientists to situate their own identities within a continuum, recognize effective techniques they may not have considered, and refine their approach based on the needs of their intended audience. This highlights the need to balance between different writing goals: too little personal language may distance readers, while too much may overshadow the science, and there is no single “right” amount that universally excites or engages audiences.

Exposing scientists to two extremes (with and without personal language) served as anchor points that enabled scientists to situate their own style choices in relation to two extreme science communication styles. This suggests that effective science communication strategies should not prescribe personal language as an all-or-nothing approach, but instead provide opportunities for scientists to choose the degree of personal language that best matches their topic, audience, and communication goals. More broadly, these findings point to the value of presenting writing guidelines as a continuum, where contrastive options can help writers reflect on and adopt strategies in ways that align with their own voice and writing context. AI-assisted writing tools can embed this principle by surfacing stylistic contrasts that guide, rather than prescribe, how these communication guidelines can be implemented in practice.

\section{Limitations and Future Work}
Our reader study only included undergraduate students in the US, and as such may not accurately reflect the general public. Additionally, we only conducted followup interviews with a subset of 8 participants which might not capture the diverse range of reader preferences. Instead, our work seeks to illuminate some reasons why readers may prefer certain strategies, and demonstrates that even within this population there are variations in reader preferences. Our dataset for science explanations only contained 15 different topics that covered 5 STEM topics (Physics, Computer Science, Civil Engineering, Psychology, and Statistics). Future work could expand the domain to life sciences and topics covered to help identify themes and trends in science communication techniques across different domains and complexities.

Our writer study only included researchers from the computer science field. Future studies should include other STEM scientists to understand how writing in different fields of study might result in different structures and style preferences. Additionally, 9 of 10 of the participants in the writer study were men which might not offer a complete understanding of expert preferences for science communication on social media. There was little age diversity in PhD students, so additional research will need to investigate how age differences affect a scientist's comfort with using different structure and style options for science communication on social media. When using the system, scientists were not asked to actually publish their final science narratives on social media. As such, scientists' choices might not fully reflect how they would have written if they were to post on social media. Finally, future studies can also evaluate the longitudinal effects of using the system to evaluate whether the novelty of seeing different structures and style options wears off with prolonged usage and whether there are any long-term benefits in seeing multiple structure and style options during the drafting process for science communication. 

\section{Conclusion}
It is crucial for science communication to engage the general public, and prior research suggests that using colloquial techniques from social media can be effective. Despite this, many scientists are hesitant to apply these techniques due to concerns about losing their authoritative voice. Our research highlights the complexity of public science communication and the need to balance readers’ and scientists’ perspectives. While readers generally preferred explanations that included examples, walkthroughs, and personal language, their preferences were nuanced and context-dependent, influenced by their personal experiences and the complexity of the topic. Conversely, scientists often resisted adopting these techniques, worried that these strategies might compromise the clarity or authority of their explanations. However, when given the opportunity to explore various narrative structure and style options, scientists were able to reflect on trade-offs, merge across options, and situate themselves along a continuum rather than treating strategies as rigid rules. These findings suggest that effective science communication support should not prescribe fixed guidelines, but instead present contrastive options that help scientists balance engagement and clarity, authority and relatability. 

\bibliographystyle{ACM-Reference-Format}
\bibliography{paper}

@misc{nielsen10usability,
  title={Usability heuristics for user interface design},
  author={Nielsen, Jakob},
  year={10}
}

@article{gigerenzer2011recognition,
  title={The recognition heuristic: A decade of research},
  author={Gigerenzer, Gerd and Goldstein, Daniel G},
  journal={Judgment and Decision Making},
  volume={6},
  number={1},
  pages={100--121},
  year={2011},
  publisher={Cambridge University Press}
}

@article{powell2008building,
  title={Building citizen capacities for participation in nanotechnology decision-making: the democratic virtues of the consensus conference model},
  author={Powell, Maria and Lee Kleinman, Daniel},
  journal={Public Understanding of Science},
  volume={17},
  number={3},
  pages={329--348},
  year={2008},
  publisher={Sage Publications Sage UK: London, England}
}

@article{burns2003science,
  title={Science communication: a contemporary definition},
  author={Burns, Terry W and O'Connor, D John and Stocklmayer, Susan M},
  journal={Public understanding of science},
  volume={12},
  number={2},
  pages={183--202},
  year={2003},
  publisher={Sage Publications}
}

@article{williams2022hci,
  title={An HCI Research Agenda for Online Science Communication},
  author={Williams, Spencer and Jones, Ridley and Reinecke, Katharina and Hsieh, Gary},
  journal={Proceedings of the ACM on Human-Computer Interaction},
  volume={6},
  number={CSCW2},
  pages={1--22},
  year={2022},
  publisher={ACM New York, NY, USA}
}

@article{gero2021makes,
  title={What makes tweetorials tick: How experts communicate complex topics on twitter},
  author={Gero, Katy Ilonka and Liu, Vivian and Huang, Sarah and Lee, Jennifer and Chilton, Lydia B},
  journal={Proceedings of the ACM on Human-computer Interaction},
  volume={5},
  number={CSCW2},
  pages={1--26},
  year={2021},
  publisher={ACM New York, NY, USA}
}

@article{schafer2017changing,
  title={How changing media structures are affecting science news coverage},
  author={Sch{\"a}fer, Mike S},
  journal={The Oxford handbook of the science of science communication},
  volume={51},
  pages={57},
  year={2017},
  publisher={Oxford University Press Oxford}
}

@article{rice2017contexts,
  title={The contexts and dynamics of science communication and language},
  author={Rice, Ronald E and Giles, Howard},
  journal={Journal of Language and Social Psychology},
  volume={36},
  number={1},
  pages={127--139},
  year={2017},
  publisher={Sage Publications Sage CA: Los Angeles, CA}
}

@inproceedings{nguyen2024simulating,
  title={Simulating climate change discussion with large language models: considerations for science communication at scale},
  author={Nguyen, Ha and Nguyen, Victoria and L{\'o}pez-Fierro, Sar{\'\i}ah and Ludovise, Sara and Santagata, Rossella},
  booktitle={Proceedings of the Eleventh ACM Conference on Learning@ Scale},
  pages={28--38},
  year={2024}
}

@inproceedings{kim2024authors,
  title={Authors' Values and Attitudes Towards AI-bridged Scalable Personalization of Creative Language Arts},
  author={Kim, Taewook and Han, Hyomin and Adar, Eytan and Kay, Matthew and Chung, John Joon Young},
  booktitle={Proceedings of the CHI Conference on Human Factors in Computing Systems},
  pages={1--16},
  year={2024}
}

@inproceedings{august2020writing,
  title={Writing strategies for science communication: Data and computational analysis},
  author={August, Tal and Kim, Lauren and Reinecke, Katharina and Smith, Noah A},
  booktitle={Proceedings of the 2020 Conference on Empirical Methods in Natural Language Processing (EMNLP)},
  pages={5327--5344},
  year={2020}
}

@article{fontaine2019communicating,
  title={Communicating science in the digital and social media ecosystem: scoping review and typology of strategies used by health scientists},
  author={Fontaine, Guillaume and Maheu-Cadotte, Marc-Andr{\'e} and Lavall{\'e}e, Andr{\'e}ane and Mailhot, Tanya and Rouleau, Genevi{\`e}ve and Bouix-Picasso, Julien and Bourbonnais, Anne and others},
  journal={JMIR public health and surveillance},
  volume={5},
  number={3},
  pages={e14447},
  year={2019},
  publisher={JMIR Publications Inc., Toronto, Canada}
}

@misc{bik2015ten,
  title={Ten simple rules for effective online outreach},
  author={Bik, Holly M and Dove, Alistair DM and Goldstein, Miriam C and Helm, Rebecca R and MacPherson, Rick and Martini, Kim and Warneke, Alexandria and McClain, Craig},
  journal={PLoS computational biology},
  volume={11},
  number={4},
  pages={e1003906},
  year={2015},
  publisher={Public Library of Science San Francisco, CA USA}
}

@article{zhang2022no,
  title={No laughing matter: Exploring the effects of scientists’ humor use on Twitter and the moderating role of superiority},
  author={Zhang, Annie L and Lu, Hang},
  journal={Science Communication},
  volume={44},
  number={4},
  pages={418--445},
  year={2022},
  publisher={SAGE Publications Sage CA: Los Angeles, CA}
}

@article{gilbert2020run,
  title={" I run the world's largest historical outreach project and it's on a cesspool of a website." Moderating a Public Scholarship Site on Reddit: A Case Study of r/AskHistorians},
  author={Gilbert, Sarah A},
  journal={Proceedings of the ACM on Human-Computer Interaction},
  volume={4},
  number={CSCW1},
  pages={1--27},
  year={2020},
  publisher={ACM New York, NY, USA}
}

@article{yuan2019should,
  title={Should scientists talk about GMOs nicely? Exploring the effects of communication styles, source expertise, and preexisting attitude},
  author={Yuan, Shupei and Ma, Wenjuan and Besley, John C},
  journal={Science Communication},
  volume={41},
  number={3},
  pages={267--290},
  year={2019},
  publisher={SAGE Publications Sage CA: Los Angeles, CA}
}

@article{yuan2020s,
  title={“It’s global warming, stupid”: Aggressive communication styles and political ideology in science blog debates about climate change},
  author={Yuan, Shupei and Lu, Hang},
  journal={Journalism \& Mass Communication Quarterly},
  volume={97},
  number={4},
  pages={1003--1025},
  year={2020},
  publisher={SAGE Publications Sage CA: Los Angeles, CA}
}

@misc{koivumaki2020social,
  title={On Social Media Science Seems to Be More Human”: Exploring Researchers as Digital Science Communicators. Media and Communication 8, 2 (2020), 425},
  author={Koivum{\"a}ki, Kaisu and Koivum{\"a}ki, Timo and Karvonen, Erkki},
  year={2020}
}

@article{oktay2014demographic,
  title={Demographic breakdown of twitter users: An analysis based on names},
  author={Oktay, Huseyin and Firat, Aykut and Ertem, Zeynep},
  journal={Academy of Science and Engineering (ASE)},
  pages={A},
  year={2014}
}

@article{hou2017hacking,
  title={Hacking with NPOs: collaborative analytics and broker roles in civic data hackathons},
  author={Hou, Youyang and Wang, Dakuo},
  journal={Proceedings of the ACM on Human-Computer Interaction},
  volume={1},
  number={CSCW},
  pages={1--16},
  year={2017},
  publisher={ACM New York, NY, USA}
}

@misc{pewresearchFactsAbout,
	author = {Katherine Schaeffer},
	title = {5 facts about how {A}mericans use {F}acebook, two decades after its launch --- pewresearch.org},
	howpublished = {\url{https://www.pewresearch.org/short-reads/2024/02/02/5-facts-about-how-americans-use-facebook-two-decades-after-its-launch/}},
	year = {2024},
	note = {[Accessed 28-01-2025]},
}

@article{hargittai2018young,
  title={How do young adults engage with science and research on social media? Some preliminary findings and an agenda for future research},
  author={Hargittai, Eszter and F{\"u}chslin, Tobias and Sch{\"a}fer, Mike S},
  journal={Social Media+ Society},
  volume={4},
  number={3},
  pages={2056305118797720},
  year={2018},
  publisher={SAGE Publications Sage UK: London, England}
}

@article{goodman_snowball_1961,
	title = {Snowball {Sampling}},
	volume = {32},
	issn = {0003-4851},
	url = {https://www.jstor.org/stable/2237615},
	number = {1},
	urldate = {2021-12-23},
	journal = {The Annals of Mathematical Statistics},
	author = {Goodman, Leo A.},
	year = {1961},
	note = {Publisher: Institute of Mathematical Statistics},
	pages = {148--170},
}

@article{bruggemann2020post,
  title={Post-normal science communication: exploring the blurring boundaries of science and journalism},
  author={Br{\"u}ggemann, Michael and L{\"o}rcher, Ines and Walter, Stefanie},
  journal={Journal of Science Communication},
  volume={19},
  number={3},
  pages={A02},
  year={2020},
  publisher={SISSA Medialab srl}
}

@article{pearson2001participation,
  title={The participation of scientists in public understanding of science activities: The policy and practice of the UK Research Councils},
  author={Pearson, Gillian},
  journal={Public Understanding of Science},
  volume={10},
  number={1},
  pages={121--137},
  year={2001},
  publisher={Sage Publications}
}

@article{national2017communicating,
  title={Communicating science effectively: A research agenda},
  author={National Academies of Sciences and Division of Behavioral and Social Sciences and Committee on the Science of Science Communication and A Research Agenda},
  year={2017},
  publisher={National Academies Press}
}

@inproceedings{lee2024design,
  title={A Design Space for Intelligent and Interactive Writing Assistants},
  author={Lee, Mina and Gero, Katy Ilonka and Chung, John Joon Young and Shum, Simon Buckingham and Raheja, Vipul and Shen, Hua and Venugopalan, Subhashini and Wambsganss, Thiemo and Zhou, David and Alghamdi, Emad A and others},
  booktitle={Proceedings of the CHI Conference on Human Factors in Computing Systems},
  pages={1--35},
  year={2024}
}

@article{10.1145/3479566,
author = {Gero, Katy Ilonka and Liu, Vivian and Huang, Sarah and Lee, Jennifer and Chilton, Lydia B.},
title = {What Makes Tweetorials Tick: How Experts Communicate Complex Topics on Twitter},
year = {2021},
issue_date = {October 2021},
publisher = {Association for Computing Machinery},
address = {New York, NY, USA},
volume = {5},
number = {CSCW2},
url = {https://doi.org/10.1145/3479566},
doi = {10.1145/3479566},
journal = {Proc. ACM Hum.-Comput. Interact.},
month = {oct},
articleno = {422},
numpages = {26}
}

@article{doi:10.1056/NEJMp1906790,
author = {Anthony C. Breu },
title = {Why Is a Cow? Curiosity, Tweetorials, and the Return to Why},
journal = {New England Journal of Medicine},
volume = {381},
number = {12},
pages = {1097-1098},
year = {2019},
doi = {10.1056/NEJMp1906790},

URL = {https://www.nejm.org/doi/full/10.1056/NEJMp1906790},
eprint = {https://www.nejm.org/doi/pdf/10.1056/NEJMp1906790}
}

@ARTICLE{Bruggemann2020-ez,
  title     = "Post-normal science communication: exploring the blurring
               boundaries of science and journalism",
  author    = "Br{\"u}ggemann, Michael and L{\"o}rcher, Ines and Walter,
               Stefanie",
  journal   = "J. Sci. Commun.",
  publisher = "Sissa Medialab Srl",
  volume    =  19,
  number    =  03,
  pages     = "A02",
  month     =  jun,
  year      =  2020
}

@article{kopke2019stepping,
  title={Stepping out of the ivory tower for ocean literacy},
  author={Kopke, Kathrin and Black, Jeffrey and Dozier, Amy},
  journal={Frontiers in Marine Science},
  volume={6},
  pages={60},
  year={2019},
  publisher={Frontiers Media SA}
}

@article{cote2018scientists,
  title={Scientists on Twitter: Preaching to the choir or singing from the rooftops?},
  author={C{\^o}t{\'e}, Isabelle M and Darling, Emily S},
  journal={Facets},
  volume={3},
  number={1},
  pages={682--694},
  year={2018},
  publisher={Canadian Science Publishing 65 Auriga Drive, Suite 203, Ottawa, ON K2E 7W6}
}

@article{sommers1980revision,
  title={Revision strategies of student writers and experienced adult writers},
  author={Sommers, Nancy},
  journal={College Composition \& Communication},
  volume={31},
  number={4},
  pages={378--388},
  year={1980},
  publisher={NCTE}
}

@inproceedings{shakeri2021saga,
  title={Saga: Collaborative storytelling with gpt-3},
  author={Shakeri, Hanieh and Neustaedter, Carman and DiPaola, Steve},
  booktitle={Companion Publication of the 2021 Conference on Computer Supported Cooperative Work and Social Computing},
  pages={163--166},
  year={2021}
}

@inproceedings{buschek2021impact,
  title={The impact of multiple parallel phrase suggestions on email input and composition behaviour of native and non-native english writers},
  author={Buschek, Daniel and Z{\"u}rn, Martin and Eiband, Malin},
  booktitle={Proceedings of the 2021 CHI Conference on Human Factors in Computing Systems},
  pages={1--13},
  year={2021}
}

@inproceedings{biermann2022tool,
  title={From tool to companion: Storywriters want AI writers to respect their personal values and writing strategies},
  author={Biermann, Oloff C and Ma, Ning F and Yoon, Dongwook},
  booktitle={Proceedings of the 2022 ACM Designing Interactive Systems Conference},
  pages={1209--1227},
  year={2022}
}

@article{lorono2018responsibility,
  title={Responsibility and science communication: scientists’ experiences of and perspectives on public communication activities},
  author={Loro{\~n}o-Leturiondo, Maria and Davies, Sarah R},
  journal={Journal of Responsible Innovation},
  volume={5},
  number={2},
  pages={170--185},
  year={2018},
  publisher={Taylor \& Francis}
}

@article{tardy2023spread,
  title={“Spread is like wildfire”: Attracting and retaining attention in COVID19 science tweetorials},
  author={Tardy, Christine M},
  journal={Ib{\'e}rica: Revista de la Asociaci{\'o}n Europea de Lenguas para Fines Espec{\'\i}ficos (AELFE)},
  number={46},
  pages={181--205},
  year={2023},
  publisher={Asociaci{\'o}n Europea de Lenguas para Fines Espec{\'\i}ficos (AELFE)}
}

@incollection{heyd2015digital,
  title={Digital genres and processes of remediation},
  author={Heyd, Theresa},
  booktitle={The Routledge handbook of language and digital communication},
  pages={87--102},
  year={2015},
  publisher={Routledge}
}

@article{wang2023improving,
  title={Improving Pacing in Long-Form Story Planning},
  author={Wang, Yichen and Yang, Kevin and Liu, Xiaoming and Klein, Dan},
  journal={arXiv preprint arXiv:2311.04459},
  year={2023}
}

@article{yang2022doc,
  title={Doc: Improving long story coherence with detailed outline control},
  author={Yang, Kevin and Klein, Dan and Peng, Nanyun and Tian, Yuandong},
  journal={arXiv preprint arXiv:2212.10077},
  year={2022}
}

@article{yang2022re3,
  title={Re3: Generating longer stories with recursive reprompting and revision},
  author={Yang, Kevin and Tian, Yuandong and Peng, Nanyun and Klein, Dan},
  journal={arXiv preprint arXiv:2210.06774},
  year={2022}
}

@article{rashkin2020plotmachines,
  title={PlotMachines: Outline-conditioned generation with dynamic plot state tracking},
  author={Rashkin, Hannah and Celikyilmaz, Asli and Choi, Yejin and Gao, Jianfeng},
  journal={arXiv preprint arXiv:2004.14967},
  year={2020}
}

@inproceedings{park2023designing,
  title={Designing a language model-based authoring tool prototype for interactive storytelling},
  author={Park, Jeongyoon and Shin, Jumin and Kim, Gayeon and Bae, Byung-Chull},
  booktitle={International Conference on Interactive Digital Storytelling},
  pages={239--245},
  year={2023},
  organization={Springer}
}

@inproceedings{belz2024story,
  title={Story-Driven: Exploring the Impact of Providing Real-time Context Information on Automated Storytelling},
  author={Belz, Jan Henry and Weilke, Lina Madlin and Winter, Anton and Hallgarten, Philipp and Rukzio, Enrico and Grosse-Puppendahl, Tobias},
  booktitle={Proceedings of the 37th Annual ACM Symposium on User Interface Software and Technology},
  pages={1--15},
  year={2024}
}

@inproceedings{calderwood2022spinning,
  title={Spinning Coherent Interactive Fiction through Foundation Model Prompts.},
  author={Calderwood, Alex and Wardrip-Fruin, Noah and Mateas, Michael},
  booktitle={ICCC},
  pages={44--53},
  year={2022}
}

@inproceedings{ding2023fluid,
  title={Fluid transformers and creative analogies: Exploring large language models’ capacity for augmenting cross-domain analogical creativity},
  author={Ding, Zijian and Srinivasan, Arvind and MacNeil, Stephen and Chan, Joel},
  booktitle={Proceedings of the 15th Conference on Creativity and Cognition},
  pages={489--505},
  year={2023}
}

@article{das2023balancing,
  title={Balancing Effect of Training Dataset Distribution of Multiple Styles for Multi-Style Text Transfer},
  author={Das, Debarati and Ma, David and Kang, Dongyeop},
  journal={arXiv preprint arXiv:2305.15582},
  year={2023}
}

@article{august2023paper,
  title={Paper plain: Making medical research papers approachable to healthcare consumers with natural language processing},
  author={August, Tal and Wang, Lucy Lu and Bragg, Jonathan and Hearst, Marti A and Head, Andrew and Lo, Kyle},
  journal={ACM Transactions on Computer-Human Interaction},
  volume={30},
  number={5},
  pages={1--38},
  year={2023},
  publisher={ACM New York, NY}
}

@misc{symplurTweetorialsFrom,
	author = {Symplur},
	title = {Tweetorials — {F}rom {E}arly {B}eginnings to {H}uge {G}rowth and {B}eyond --- symplur.com},
	howpublished = {\url{https://www.symplur.com/blog/tweetorials-from-early-beginnings-to-huge-growth-and-beyond/}},
	year = {2019},
	note = {[Accessed 13-01-2025]},
}

@inproceedings{hullman2018improving,
  title={Improving comprehension of measurements using concrete re-expression strategies},
  author={Hullman, Jessica and Kim, Yea-Seul and Nguyen, Francis and Speers, Lauren and Agrawala, Maneesh},
  booktitle={Proceedings of the 2018 CHI Conference on Human Factors in Computing Systems},
  pages={1--12},
  year={2018}
}

@inproceedings{10.1145/3643834.3661587,
author = {Long, Tao and Gero, Katy Ilonka and Chilton, Lydia B},
title = {Not Just Novelty: A Longitudinal Study on Utility and Customization of an AI Workflow},
year = {2024},
isbn = {9798400705830},
publisher = {Association for Computing Machinery},
address = {New York, NY, USA},
url = {https://doi.org/10.1145/3643834.3661587},
doi = {10.1145/3643834.3661587},
booktitle = {Proceedings of the 2024 ACM Designing Interactive Systems Conference},
pages = {782–803},
numpages = {22}
}

@article{della2021expert,
  title={Expert communication on Twitter: Comparing economists’ and scientists’ social networks, topics and communicative styles},
  author={Della Giusta, Marina and Jaworska, Sylvia and Vukadinovi{\'c} Greetham, Danica},
  journal={Public understanding of science},
  volume={30},
  number={1},
  pages={75--90},
  year={2021},
  publisher={SAGE Publications Sage UK: London, England}
}

@misc{long2023tweetorialhooksgenerativeai,
      title={Tweetorial Hooks: Generative AI Tools to Motivate Science on Social Media}, 
      author={Tao Long and Dorothy Zhang and Grace Li and Batool Taraif and Samia Menon and Kynnedy Simone Smith and Sitong Wang and Katy Ilonka Gero and Lydia B. Chilton},
      year={2023},
      eprint={2305.12265},
      archivePrefix={arXiv},
      primaryClass={cs.HC},
      url={https://arxiv.org/abs/2305.12265}, 
}

@book{charmaz_constructing_2006,
	address = {London ; Thousand Oaks, Calif},
	author = {Charmaz, Kathy},
	isbn = {978-0-7619-7352-2 978-0-7619-7353-9},
	language = {en},
	publisher = {Sage Publications},
	title = {Constructing grounded theory},
	year = {2006}}

@book{merriam2015qualitative,
  title={Qualitative research: A guide to design and implementation},
  author={Merriam, Sharan B and Tisdell, Elizabeth J},
  year={2015},
  publisher={John Wiley \& Sons}
}

@article{holtzblatt2017affinity,
  title={The affinity diagram},
  author={Holtzblatt, Karen and Beyer, Hugh},
  journal={Contextual design},
  pages={127--146},
  year={2017},
  publisher={Elsevier}
}

%TC:ignore
\appendix
\newpage
\section{STEM Topics}
\label{stem_topics}

\textbf{Computer Science }: \\
(Introductory)  Depth-First Search\\
(Intermediate)  Back Propagation \\
(Advanced)      Recurrent Neural Networks\\

\noindent\textbf{Physics}: \\
(Introductory)  Distance and Displacement\\
(Intermediate)  Thermal Equilibrium \\
(Advanced)      Thin Film Interference\\

\noindent \textbf{Statistics}: \\
(Introductory)  Normal Distribution\\
(Intermediate)  Central Limit Theorem \\
(Advanced)      Linear Regression\\

\noindent \textbf{Civil Engineering}: \\
(Introductory)  Lattice Structure\\
(Intermediate)  Tensile Structure \\
(Advanced)      Curtain Wall System\\

\noindent \textbf{Psychology}: \\
(Introductory)  Retroactive and Proactive Interference\\
(Intermediate)  Feature Integration Theory \\
(Advanced)      Walker's Action Decrement Theory\\

\section{GPT Generation Prompts}
\label{prompts}

\subsection{EWP}

[FewShot] \footnote{EWP-NoFewShot does not have this part.}\\

\noindent Instructions:\\
Talk to a friend about the topic: [topic] in the domain: [domain].\\
Explain how the topic works.\\
Use the given example to help explain how the topic words: [example].\\
 Use the scenario to provide additional context: [scenario].\\

\noindent Output format: \\
a piece of writing with short paragraphs (280 characters for each paragraph).\\

\noindent Walkthrough Guidelines:\\
Tell the story from a first-person perspective.\\
Walk through the story timeline in a sequence. \\
Be sure to explain each dimension of the topic in detail, relating it back to the given example and scenario.\\

\noindent Emotional Guidelines:\\
Take the second-person audience on an emotional journey.\\
Add visually descriptive details in the storytelling.\\
Use emotional languages (both negative and positive).\\
Add questions that echo with the audience and spark curiosity.

\subsection{WP}
\label{WP_prompt}

Narrative: [Output from EWP]\\

\noindent Instructions:\\
You are given a science narrative that explains how [topic] works.\\
Keep the same tone and structure as the given narrative.\\
Remove the example of [example\_label] from the narrative.\\
Do not include ANY examples.\\
Only provide a technical walkthrough of [topic] following the same structure.

\subsection{EP-NoFewShot}
\label{EP_prompt}

Instructions:\\
Write a series of Tweets explaining the given topic: [topic] in the domain of [domain].\\
Make sure each Tweet is less than 280 characters.\\

\noindent Do not use technical jargon and define all technical components.\\
Do not walkthrough using timeline sequence. \\
Do not use words such as ``before", ``after", ``then", ``next", ``also", ``first", ``second", ``third", ``last", ``summary".\\
Explain a different technical component of the topic in each tweet in a non-sequential modular approach.\\
Each tweet stands alone and allows the reader to navigate through the explanations in various orders.\\

\subsection{EW}
\label{EW_prompt}

Narrative: [Output from EWP]\\

\noindent Instructions:\\
You are given a science narrative that uses emotional and engaging language to explain a concept.\\
Your task is to write a new narrative that maintains the same structure of the given narrative but removes all emotional language and all subjective language.\\
Edit the sentences to remove `you', `I', and `we' pronouns. \\
Do not use rhetorical or confirmation questions.\\
Maintain the active voice and use ``people" or ``student" or other general terms as the subject.\\
Use objective and generalizable language wherever possible.\\
Remove any extraneous descriptions and adjectives.\\
Use formal language.\\
%TC:endignore

\end{document}